\documentclass[prb,twocolumn,superscriptaddress,showpacs,preprintnumbers,amsmath,amssymb]{revtex4}

\usepackage{graphicx}
\usepackage{dcolumn}
\usepackage{bm}
\usepackage{enumerate}

\newcommand{\Tr}{\rm Tr}

\begin{document}

\bibliographystyle{apsrev} 

\title{Inhomogeneous dynamic nuclear polarization and suppression of electron-polarization decay in a quantum dot}

\author{Na Wu}
\affiliation{School of Physics and Technology, Wuhan University, Wuhan, Hubei 430072, China}
\affiliation{Department of Optical Science and Engineering, Fudan University, Shanghai 200433, China}

\author{Wenkui Ding}
\affiliation{School of Physics and Technology, Wuhan University, Wuhan, Hubei 430072, China}

\author{Anqi Shi}
\affiliation{School of Physics and Technology, Wuhan University, Wuhan, Hubei 430072, China}

\author{Wenxian Zhang}
\email{wxzhang@whu.edu.cn}
\affiliation{School of Physics and Technology, Wuhan University, Wuhan, Hubei 430072, China}

\date{\today}

\begin{abstract}
We investigate the dynamic nuclear polarization process by frequently injecting polarized electron spins into a quantum dot. Due to the suppression of the direct dipolar and indirect electron-mediated nuclear spin interactions, by the frequently injected electron spins, the analytical predictions under the independent spin approximation agree well with quantum numerical simulations. Our results show that the acquired nuclear polarization is highly inhomogeneous, proportional to the square of the local electron-nuclear hyperfine interaction constant, if the injection frequency is high. Utilizing the inhomogeneously polarized nuclear spins as an initial state, we further show that the electron-polarization decay time can be extended 100 times even at a relatively low nuclear polarization ($<20\%$), without much suppression of the fluctuation of the Overhauser field. Our results lay the foundation for future investigations of the effect of DNP in more complex spin systems, such as double quantum dots and nitrogen vacancy centers in diamonds.
\end{abstract}


\maketitle

\section{Introduction}

Quantum coherence is of key importance for practical quantum devices, such as quantum computers~\cite{Nielsen00, Ladd10} and spintronic devices~\cite{Zutic04}. To maintain the coherence of a quantum system, there are basically two kinds of methods: (I) Effectively reducing the system and environment coupling strength such as dynamical decoupling~\cite{Slichter92, Viola99} and employing decoherence free subspace~\cite{Viola00, Khodjasteh05}; (II) Engineering the environment to a certain states, e.g., a dark state, a many-body singlets, which only weakly affect the system dynamics~\cite{Scully97, Stepanenko06, Ribeiro09, Yao11}.

Electron spins in semiconductor quantum dots (QDs) are promising candidate of solid-state quantum bits (qubits), due to their long relaxation time in strong magnetic fields~\cite{Loss98, Kane98, Zutic04}. However, the QD electron spins, which are decohered by surrounding nuclear spins through hyperfine coupling, have a rather short coherence/relaxation time, typically in the order of 10 ns in a few mT magnetic field and at $\sim$100 mK low
temperature~\cite{Petta05,Koppens05,Johnson05,Merkulov02,Zhang06,Deng06}. Thus,  extension of electron spin coherence time is highly
desired~\cite{Greilich07, Reilly08, Xu09}.

Many methods have been proposed to prolong the coherence time of the electron spins, including dynamical decoupling via coherent
control pulses~\cite{Petta05, Yao07, Witzel07, Zhang07a, Khodjasteh07, Zhang08} and quantum environment engineering (nuclear spin state narrowing and nuclear spin polarization)~\cite{Klauser06, Stepanenko06, Ramon07, Reilly08, Ribeiro09, Yao11}. For the dynamical decoupling and nuclear spin
state narrowing, theoretical results show that the coherence time can be
extended more than 2 orders of magnitude~\cite{Zhang07a, Zhang07b,
Stepanenko06}, but the experimental confirmation is still illusive, due to the challenging requirement of the high repetition rate of the control pulses or the measurements.

Nuclear spin polarization has been extensively investigated both theoretically and experimentally. In general, the nuclear spins can be polarized by applying a
very strong but static magnetic field (usually above 1 Tesla)~\cite{Slichter92, Burkard99, Coish04, Deng06} or by repeatedly injecting polarized electron spins with optical or electrical pumping methods~\cite{Brown96, Bracker05, Ono02, Reilly08, Petta08, Xu09, Reilly10, Issler10, Sun12}. We refer the former as static nuclear polarization and the latter as dynamic nuclear polarization (DNP). For the static nuclear polarization, the electron spin coherence time is almost unchanged in a QD with uniformly polarized nuclear spins, unless the nuclear polarization is above 90\% which is unexpectedly high in experiments~\cite{Burkard99, Coish04, Zhang06, Deng06}. For the DNP, early experiments and theories show that the coherence time of the electron spin in double QDs can be extended significantly by suppressing the fluctuation of the Overhauser field difference~\cite{Ramon07, Reilly08, Ribeiro09, Barthel12}, but subsequent experiments and more recent theories support a different interpretation~\cite{Foletti09, Gullans10, Bluhm10, Zhang10b, Gullans13}. A complete physical understanding of the DNP process, rooted in a microscopic picture, is still lacking.

In this paper, we focus on the DNP process in a QD where the hyperfine coupling between the central electron spin and the surrounding nuclear spins is dominant. We develop analytical equations for the DNP dynamics under the independent spin approximation (ISA). Numerical quantum simulations with small-scale nuclear spin bath are further performed to confirm the validity of the analytical predictions. We ascribe the success of ISA to the suppression of the direct dipolar and indirect electron-mediated nuclear spin interactions by frequently injected electron spins. Furthermore, we find that the nuclear polarization acquired through DNP processes is highly inhomogeneous: strongly coupled nuclear spins gain much more polarization. Consequently, the electron-polarization decay is suppressed significantly, if the final DNP state is employed as an initial nuclear spin state, due to the protection effect by these highly polarized nuclear spins, instead of the suppression of the Overhauser field fluctuation.

The paper is organized as follows. We describe the QD system in Sec.~\ref{sec:sys}. The dynamics of a DNP process for a single nuclear spin and many nuclear spins are presented in Sec.~\ref{sec:dnp}. In Secs.~\ref{sec:dnpB} and~\ref{sec:dnpD} we discuss respectively the effect of external magnetic fields and the dipolar coupling between nuclear spins during a DNP process. In Sec.~\ref{sec:dnpS} the suppression of the electron polarization decay is shown with the initial nuclear spin state being an inhomogeneously polarized one. Finally, the conclusion is given in Sec.~\ref{sec:con}.

\section{Spin system in a QD}
\label{sec:sys}

We consider a singly charged quantum dot where the electron spin couples with the surrounding nuclear spins via Fermi contact hyperfine coupling
\begin{eqnarray}
{\cal H} &=& \omega_0 S_z + {\bf S} \cdot \sum_{k=1}^N A_k {\bf I}_k + \sum_{k=1}^N \sum_{j\neq k}  \Gamma_{jk} \left(3I_{kz}I_{jz} - {\bf I}_k \cdot {\bf I}_j\right), \nonumber \\
\label{eq:h}
\end{eqnarray}
where $\omega_0$ is the electron Zeeman splitting in an external magnetic field along $z$ direction, ${\bf S}$ ($S=1/2$) is the electron spin, and ${\bf I_k}$ ($I_k=1/2, k=1,2,\cdots, N$ with $N$ the total number of nuclear spins) is the $k$th nuclear spin~\cite{Zhang06, Taylor07, Merkulov02}. For an electron in a QD, $A_k = (8\pi/3)g_e^* \mu_B g_n \mu_n |\phi({\bf x}_k)|^2$ is the Fermi contact hyperfine coupling constant, which is proportional to $|\phi({\bf x}_k)|^2$, the electron density at the $k$th nucleus. Here $g_e^*$ and $g_n$ are the Land\'e factors of the electron and the nuclei, respectively. The dipolar coupling strength between nuclear spins $j$ and $k$ are characterized by $\Gamma_{jk}$. We set $\hbar=1$ for simplicity. The Zeeman terms of nuclear spins have been removed by adopting a rotating reference frame with the same frequency as that of the nuclear Larmor precession.

For a gate-defined GaAs QD~\cite{Taylor07}, the number of nuclear spins is about $10^4 \sim 10^6$. The Zeeman splitting $\omega_0$ for the electron spin ranges from 0 to 200 $\mu$eV, proportional to the external magnetic field. Due to the inhomogeneity of $|\phi|^2$, the hyperfine coupling constant $A_k$ is in general non-uniform and is typically in the order of $A_k \gtrsim 0.1$ neV. The dipolar coupling strength (nearest neighbor) is about $\Gamma_{jk} \sim 0.01$ neV, much smaller than $A_k$. We neglect the small nuclear dipolar interaction, i.e., $\Gamma_{jk} = 0$, unless otherwise stated.

\section{DNP}
\label{sec:dnp}

\begin{figure}
\includegraphics[width=3.25in]{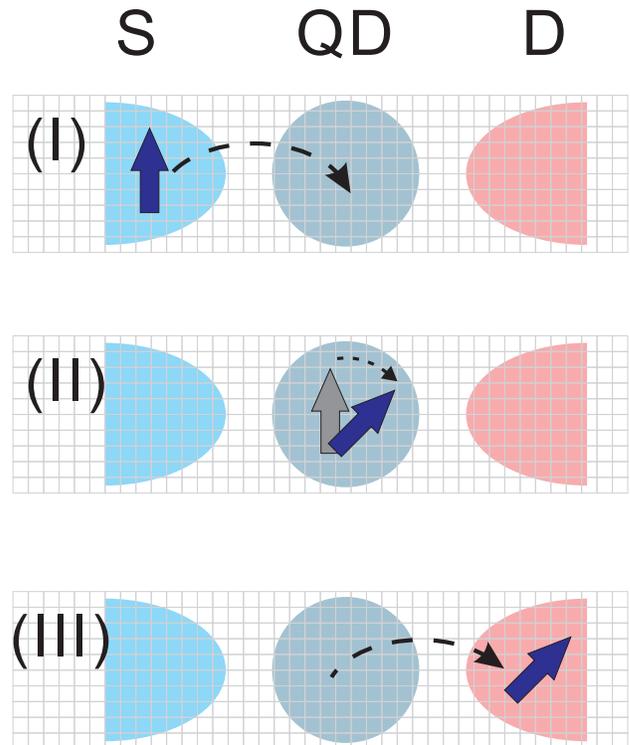}
\caption{\label{fig:dnp} (Color online) Schemetic of DNP process in a QD. S denotes the source of polarized electron spins (blue arrow), QD the quantum dot, and D the drain of the electron. (I) Injection of a polarized electron spin into the QD. (II) Due to the nonuniform hyperfine interaction between the electron and the nuclei in the QD, the electron polarization is inhomogeneously transferred to the nuclear spins. (III) Ejection of the depolarized electron. After a DNP cycle, the total nuclear polarization is increased. The polarization of the strongly interacting nuclear spins with the electron increases faster than that of the weakly interacting ones.}
\end{figure}

The DNP process is consisted of many cycles. A typical DNP cycle can in principle be divided into approximately three steps as shown in Fig.~\ref{fig:dnp}:
\begin{enumerate}[(I)]
  \item Inject a polarized electron spin into the QD. The density matrix of the total system becomes
      \begin{equation}
      \rho_0 = \rho_0^e \otimes \rho_0^n,
      \end{equation}
      where $\rho_0^e = |\uparrow\rangle \langle \uparrow|$ denotes the polarized electron spin state and $\rho_0^n$ denotes the nuclear spin state.
  \item Let the coupled spin system mix and the electron polarization is transferred to the nuclear spins, due to the total polarization conservation of the hyperfine coupling. The evolved density matrix after a cycle time $\tau$ is
      \begin{eqnarray}
      \rho(\tau) &=& |\uparrow\rangle \langle \uparrow| \otimes \rho_{\uparrow\uparrow}^n + |\uparrow\rangle \langle \downarrow| \otimes \rho_{\uparrow\downarrow}^n \nonumber \\
      && + |\downarrow\rangle \langle \uparrow| \otimes \rho_{\downarrow\uparrow}^n + |\downarrow\rangle \langle \downarrow| \otimes \rho_{\downarrow\downarrow}^n.
      \end{eqnarray}

  \item Eject the electron out of the QD~\cite{Dobrovitski06}. The nuclear spin state after a DNP cycle then becomes
      \begin{eqnarray}
      \rho^n(\tau) &=& \langle \uparrow | \rho(\tau) | \uparrow \rangle + \langle \downarrow | \rho(\tau) | \downarrow \rangle \nonumber \\
      &=& \rho_{\uparrow\uparrow}^n + \rho_{\downarrow\downarrow}^n.
      \end{eqnarray}
\end{enumerate}

Repeat the steps (I)---(III) till the nuclear spins reach a required polarization. In experiments~\cite{Reilly08, Reilly10}, the DNP cycle period $\tau$ can be as small as 250 ns, corresponding to a frequency of 4 MHz.

\subsection{DNP of a nuclear spin}

As a starting point, we consider a single electron spin coupled to a nuclear spin. The Hamiltonian without a magnetic field is simplified as
\begin{eqnarray}
{\cal H} &=& A {\bf S} \cdot {\bf I}.
\end{eqnarray}
The injected electron is fully polarized with a polarization $p_s \equiv 2\langle S_z \rangle = 1$. The initial nuclear polarization is $p \equiv 2\langle I_z \rangle$. The initial state of the total system is given by
$$\rho_0 = |\uparrow\rangle \langle \uparrow| \otimes \left({1\over 2} + pI_z \right).$$
It is easy to analytically find that the evolution of the electron polarization is~\cite{Zhang06}
\begin{equation}
p_s(t) = \frac{1+p}2 + \frac{1-p} 2 \cos(At).
\end{equation}
Due to the conservation of the total polarization, the polarization gained by the nuclear spin after a DNP cycle with a cycle period $\tau$ is
\begin{eqnarray}
\Delta p &=& 1 - p_s \nonumber \\
    &=& \frac{1-p}2\, [1-\cos(A\tau)].
\end{eqnarray}
If $\Delta p \ll 1$ for a single DNP cycle, or equivalently, $A \tau \ll 1$, the nuclear polarization difference $\Delta p$ in the above equation can be approximated by a differential $dp$. We then obtain the following differential equation for the nuclear spin polarization during the DNP process
\begin{eqnarray}
\frac{dp}{dn} \approx \frac{A^2\tau^2}4 (1-p),
\label{eq:sdnp}
\end{eqnarray}
where $n$ denotes the number of DNP cycle. Note that we have also taken the approximation $\cos (A\tau) \approx 1-(A^2\tau^2/2)$ since $A\tau \ll 1$.

\begin{figure}
\includegraphics[width=3.25in]{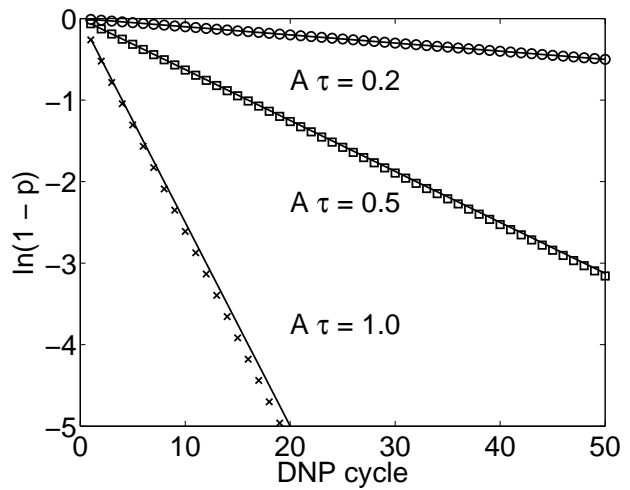}
\caption{\label{fig:single} DNP of a nuclear spin with $A\,\tau = 0.2$ (circles), $0.5$ (squares), and $1.0$ (crosses). The solid lines are obtained from Eq.~(\ref{eq:single}).}
\end{figure}

Assuming the nuclear spin is initially unpolarized and solving the equation~(\ref{eq:sdnp}), we obtain the nuclear polarization after $n$ cycles is
\begin{equation}
\label{eq:single}
p(t=n \tau) = 1-e^{-\frac{A^2\tau^2} 4 n}.
\end{equation}
It is clearly shown in Fig.~\ref{fig:single} that the nuclear polarization exponentially approaches its saturation value of 1. At the initial stage of the DNP process, where $A^2\tau^2n/4 \ll 1$, the nuclear polarization can be well approximated by $p\approx A^2\tau^2n/4$, which indicates that the nuclear polarization is proportional to the square of the coupling strength. At the nearly saturated stage, where $A^2\tau^2n/4 \gg 1$, we may approximate the nuclear polarization as $p\approx \tanh(A^2\tau^2n/8)$, as previously used~\cite{Zhang06, Zhang10b}. We remark that the analytical prediction deviates from the exact numerical calculation once the condition $A\tau \ll 1$ is seriously violated, as shown in the case of $A\tau = 1.0$ in Fig.~\ref{fig:single}.

If $\Delta p \ll 1$ is not satisfied, we get the following recursion relation between the nuclear polarizations of the $(n-1)$th cycle and the $n$th cycle
\begin{equation}
p_{n} = p_{n-1} + \frac{1-p_{n-1}} 2 \;[1-\cos(A\tau)].
\end{equation}
It is straightforward to find that
\begin{equation}
p_n = 1-\left[\frac{1+\cos(A\tau)} 2\right]^n,
\end{equation}
where $p_0=0$ has been adopted. This is an exact solution. In the large $n$ limit, the nuclear polarization becomes
\begin{equation}
p_n = 1-e^{-[1-\cos(A\tau)]n/2}.
\end{equation}
It is easy to obtain Eq.~(\ref{eq:single}) by further assuming $A\tau \ll 1$.

\subsection{DNP of a few nuclear spins}

\begin{figure}
\includegraphics[width=3.25in]{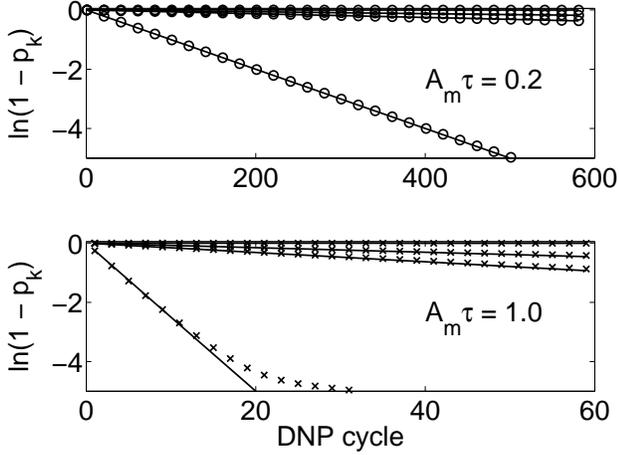}
\caption{\label{fig:many4} DNP of $N=4$ nuclear spins with $A_m\tau = 0.2$ (top) and $1.0$ (bottom), where $A_m$ is the largest coupling constant. The solid lines are obtained from Eq.~(\ref{eq:single}).  The agreement between numerical and analytical results shows that the ISA is valid if $A\,\tau$ is small.}
\end{figure}
\begin{figure}
\includegraphics[width=3.25in]{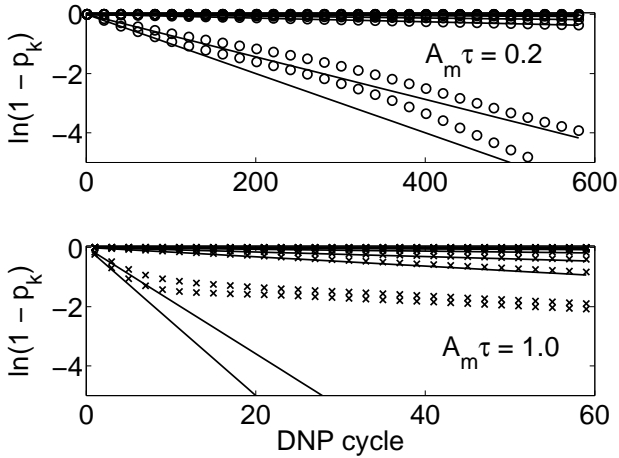}
\caption{\label{fig:many12} The same as in Fig.~\ref{fig:many4} except that the number of nuclear spins is $N=12$.}
\end{figure}

It is challenging to exactly describe the DNP dynamics for many nuclear spins. By assuming the same coupling constant for all the nuclear spins, the DNP process in double quantum dots has been numerically investigated~\cite{Ramon07, Gullans10, Gullans13}. While for different hyperfine coupling constant, it is only possible to numerically investigate the DNP process for a few nuclear spins $N \lesssim 15$.

We consider a few nuclear spins ($N=4,12$) coupled to the electron spin with Gaussian distributed random coupling constant $A_k$. The DNP process is investigated by numerically solving the von Neumann equation for the total density matrix
\begin{equation}
i\hbar \frac{\partial \rho(t)} {\partial t} = [H, \rho(t)],
\end{equation}
during the spin transfer time $t\in (n\tau,[n+1]\tau)$. The injection of the polarized electron spin at $t=n\tau$ is
\begin{equation}
\rho(t=n\tau) = |\uparrow\rangle \langle \uparrow | \otimes \rho_n(n\tau).
\end{equation}
The ejection at the end of the DNP cycle $t=(n+1)\tau$ is a partial trace over the electron spin
\begin{eqnarray}
\rho_n(t) &\equiv & {\rm Tr}_S\{\rho(t)\} \nonumber \\
    &=& \langle \uparrow| \rho(t) | \uparrow \rangle + \langle \downarrow| \rho(t) | \downarrow \rangle.
\end{eqnarray}

Since the interactions among the nuclear spins are much smaller than the hyperfine interaction between the electron and the nuclear spins, we may neglect the nuclear spin interactions, including the direct dipolar interaction and the indirect electron-mediated-nuclear-spin interaction, to the leading order. Such an approximation is hereafter referred as independent spin approximation (ISA). Under the ISA, each nuclear spin is polarized independently by the electron spin so that the nuclear polarization obeys the analytical prediction of a single spin [see Eq.~(\ref{eq:single})]: $p_k(t=n\tau) = 1-\exp(-A_k^2 \tau^2 n/4)$. Obviously, the ISA is valid if the the effect of the nuclear dipolar interaction and the indirect electron-mediated-nuclear-spin interaction is small.

During the DNP process, we quantitatively investigate the polarization of the $k$th nuclear spin $p_k$ for different DNP cycle time $\tau$. The evolution of nuclear spin polarization during DNP process, as well as the analytical predictions from Eq.~(\ref{eq:single}), are presented in Fig.~\ref{fig:many4} and~\ref{fig:many12} for $N=4$ and $12$, respectively. For both $N$s, we choose two typical values of $\tau$: (i) $A_m \tau = 0.2$ is in the small $\tau$ region where the ISA is valid; (ii) $A_m \tau = 1.0$ is in the large $\tau$ region where the ISA is violated, especially after many DNP cycles.

By comparing the numerical results with the ISA predictions in both Figs.~\ref{fig:many4} and~\ref{fig:many12}, we find good agreement at small $A_m \tau$ ($A_m$ is the largest coupling constant), indicating the validity of the ISA for a few nuclear spins. The good agreement between the ISA predictions and the numerical results implies that fast DNP cycles effectively decouple the indirect electron-mediated-nuclear-spin interaction. This decoupling effect becomes more prominent if the DNP cycle period is small, similar to the quantum Zeno effect~\cite{Misra77, Itano90, Klauser08}. Given typical experimental conditions, where $A_k \sim 0.1$ neV and $\tau \sim 50$ ns~\cite{Reilly08}, we obtain $A_k \tau \sim 0.001$ which is safely in the valid region of ISA. While for large $A_m \tau$, the ISA predictions deviate from the numerical results. The more the nuclear spins are, the larger the deviation is, as shown in Fig.~\ref{fig:many4} and~\ref{fig:many12}.

Figures~\ref{fig:many4} and~\ref{fig:many12} also show that the nuclear polarization is inhomogeneous: strongly coupled nuclear spins are polarized much faster than the weakly coupled ones. The polarization of these strongly coupled spins reaches nearly 100\% while the polarization of those weakly coupled ones is still very small. These results agree quantitatively with previous predictions of the two-region QD model~\cite{Zhang10b}.

\begin{figure}
\includegraphics[width=3.25in]{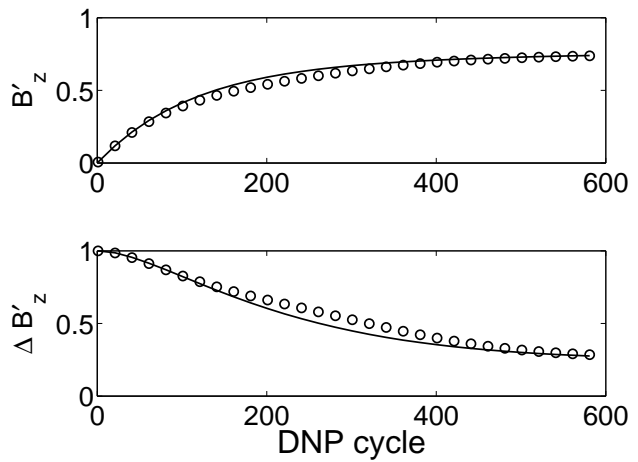}
\caption{\label{fig:12BDB} The reduced Overhauser field (top) and the reduced fluctuation (bottom) during a DNP process for $N=12$ nuclear spins with $A_m\tau = 0.2$. The solid lines are calculated under the ISA with Eqs.~(\ref{eq:single}), (\ref{eq:ismb}), and (\ref{eq:ismr}). The coupling constant $A_k$ is the same as that in Fig.~\ref{fig:many12}.}
\end{figure}

With current experimental techniques, it is a big challenge to measure directly the polarization of each nuclear spin, in particular, spatially inhomogeneous nuclear polarization, during the DNP process. But the Overhauser field $\mathbf B$ induced by the nuclear spin polarization and the fluctuation of the Overhauser field $\Delta \mathbf B$ are reachable~\cite{Reilly08, Bluhm10}. Therefore, we quantitatively compute these properties for different DNP cycle time $\tau$, as shown in Fig.~\ref{fig:12BDB} for $N=12$ nuclear spins.

In general, the Overhauser field is a vector, $\mathbf B = \sum_{\alpha=x,y,z} B_\alpha  \mathbf{e}_\alpha$ and its fluctuation along an $\alpha$-axes is $\Delta B_\alpha = \sqrt{\langle \hat B_\alpha^2 \rangle - B_\alpha^2}$, where
\begin{eqnarray}
B_\alpha &\equiv & \langle \hat B_\alpha \rangle = \left\langle \sum_k A_k \hat I_{k,\alpha} \right \rangle, \nonumber \\
\langle \hat B_\alpha^2 \rangle &=& \left\langle \sum_k \sum_\ell A_k A_\ell \hat I_{k,\alpha} \hat I_{\ell, \alpha} \right\rangle. \nonumber
\end{eqnarray}
It is easy to check that if the initial nuclear spin state is unpolarized, then during the whole DNP process $\langle I_{kx}\rangle = \langle I_{ky}\rangle = 0$. Therefore, $B_x = B_y = 0$, and $\Delta B_x = \Delta B_y =(1/2)\sqrt{\sum_k A_k^2}\,$, which are constant of motion. While along the $z$-direction, $B_z$ and $\Delta B_z$ are both time dependent, as shown in Fig.~\ref{fig:12BDB}. Under the ISA, it is straightforward to calculate that (see Appendix~\ref{sec:appd})
\begin{eqnarray}
\label{eq:ismb}
B_z &=& \frac 1 2 \sum_k A_k p_k\,,\nonumber \\
\Delta B_z &=& \frac 1 2 \sqrt{\sum_k A_k^2 (1-p_k^2)}\quad,
\end{eqnarray}
where $p_k=1-\exp(-A_k^2 \tau^2 n/4)$ is given by Eq.~(\ref{eq:single}).

We present in Fig.~\ref{fig:12BDB} actually the reduced quantities, which are normalized to their maximal value
\begin{eqnarray}
\label{eq:ismr}
B'_z &=& \frac {2B_z} {\sum_k A_k }\,,\nonumber \\
\Delta B'_z &=& \frac {2\Delta B_z} {\sqrt{\sum_k A_k^2 }}\,.
\end{eqnarray}
Once again, we find from Fig.~\ref{fig:12BDB} that the analytical ISA results are pretty close to the numerical results, indicating that the ISA is a good approximation for the parameter of $A_m\tau=0.2$. For smaller $A_m \tau$ as in experiments, one would expect much better agreement between the analytical and the numerical results.

The rapid increase of $B'_z$ at the small number of DNP cycles in Fig.~\ref{fig:12BDB} implies that the strongly coupled nuclear spins are firstly polarized. Under the ISA, the increase rate is proportional to $A_k^3$ and to the number of DNP cycles $n$. Once these strongly coupled nuclear spins are nearly fully polarized, the increase of $B'_z$ becomes slow, as shown at the large number of DNP cycles, due to the smallness of the hyperfine constant $A_k$ of the weakly coupled nuclear spins. Quite differently, the fluctuation shows slow decrease when the Overhauser field is small but rapid decrease when the strongly coupled nuclear spins are approaching the fully polarized state. The reason lies in that the fluctuation of the field is roughly proportional to $\sqrt{1-p_k^2}$ for these strongly coupled nuclear spins, which is approximately a constant (zero decrease rate) at small number of DNP cycles [$n\lesssim 50$ in the bottom panel of Fig.~\ref{fig:12BDB}] since $p_k$ is small. Once the strongly coupled nuclear spins are nearly fully polarized, i.e. $p_k \sim 1$ for some nuclear spins, the fluctuation decreases significantly. However, we notice that the fluctuation of the Overhauser field is only suppressed around 1/3 of its initial value even at $B'_z \approx 0.8$, which implies that the DNP is not an efficient way to suppress the fluctuation in a QD.

\subsection{DNP in a QD}

\begin{figure}
\includegraphics[width=3.25in]{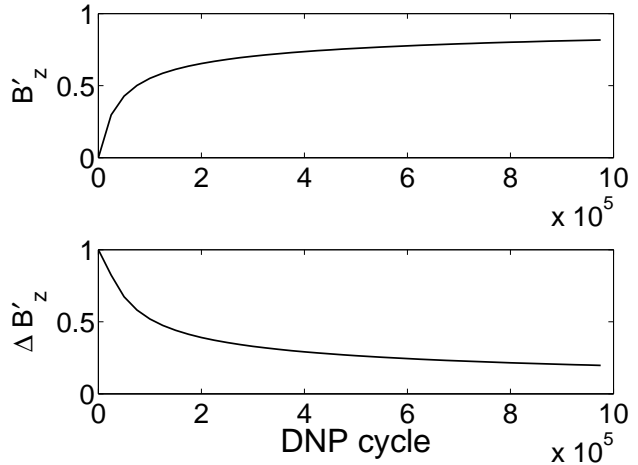}
\caption{\label{fig:BDB} The reduced Overhauser field (top) and the reduced  fluctuation (bottom) during DNP process for $N=10^6$ nuclear spins with $A_m\tau = 0.02$. The results are predictions under ISA with Eqs.~(\ref{eq:single}), (\ref{eq:ismb}), and (\ref{eq:ismr}). The coupling constant $A_k$ is randomly distributed with a Gaussian distribution function.}
\end{figure}

For a gate-defined GaAs QD~\cite{Loss98, Petta05, Johnson05, Koppens05, Petta08, Taylor07}, the number of nuclear spins is in the order of $10^6$ and the hyperfine coupling is spatially inhomogeneous. Directly simulating the DNP process without approximation in such a huge system is far beyond the capability of current computers. Instead, we adopt the ISA which has been proved to be a good approximation at small number of nuclear spins and at small $A_m \tau$.

The distribution of the coupling constants $A_k$ is assumed in a Gaussian form
\begin{equation}
A_k = A_0 \exp\left(-\frac{x_k^2+y_k^2}{\sigma_\perp^2} - \frac{z_k^2}{\sigma_z^2} \right),
\end{equation}
where $A_0=1$ is the maximal coupling constant at the center, $x_k, y_k$ are random number in the region [-1, 1], and $z_k$ is random number in the region [-0.1, 0.1]. The position of the $k$th nuclear spin is given by ($x_k, y_k, z_k$). We set $\sigma_\perp = 0.2$ and $\sigma_z = 0.1\times\sigma_\perp = 0.02$, which describes a pancake shape QD.

We plot the analytical results under ISA in Fig.~\ref{fig:BDB} for $N=10^6$ and $A_m\tau = 0.02$. Similar to the case of $N=12$, the Overhauser field increases quickly at small $B'_z$ and becomes slow once $B'_z$ is above 0.5. The Overhauser field fluctuation decreases quickly at the middle region of the DNP process, around $n\sim 2\times 10^4$. Furthermore, the suppressed fluctuation is about 1/4 at $B'_z$ being roughly 0.9, which shows that the DNP method is not good at suppressing the Overhauser field fluctuation.

\section{Effect of external magnetic fields}
\label{sec:dnpB}

\begin{figure}
\includegraphics[width=3.25in]{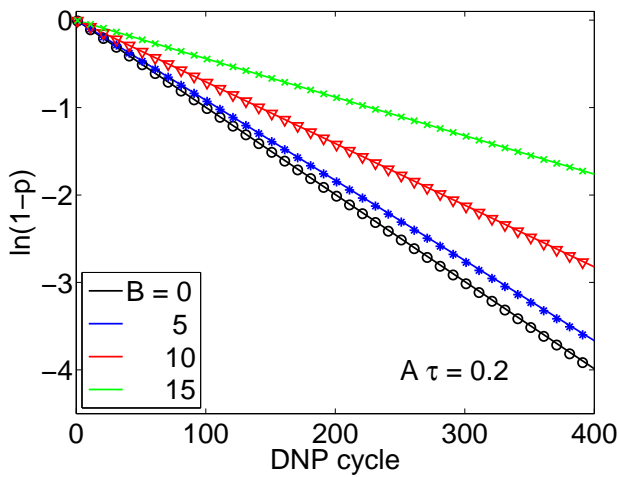}
\caption{\label{fig:B1} (Color online) The effect of external magnetic fields during the DNP process of a single nuclear spin. The fields are $\omega_0=0$ (black circles), $\omega_0=5$ (blue asterisks), $\omega_0=10$ (red triangles), and $\omega_0=15$ (green crosses), in units of $A$. The parameters are $A \tau = 0.2$. The solid lines are calculated from the analytical expression Eq.~(\ref{eq:B}). The nuclear polarization increases slower in a larger magnetic field.}
\end{figure}
\begin{figure}
\includegraphics[width=3.25in]{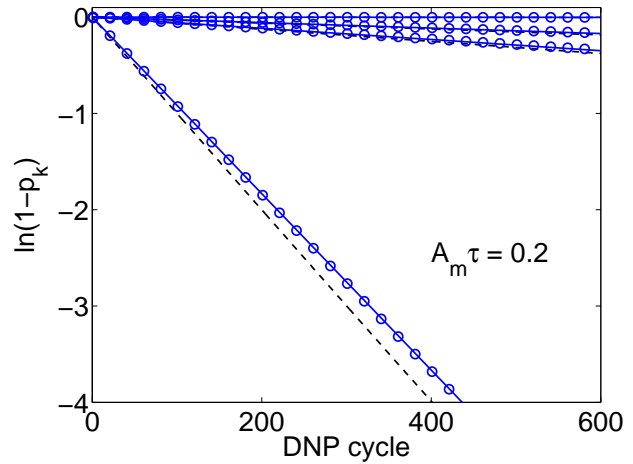}
\caption{\label{fig:B4} (Color online) The same as Fig.~\ref{fig:B1} except the nuclear spin number is $N=4$. The black dashed lines are the nuclear polarizations of the 4 spins for $\omega_0=0$, the blue circles are for $\omega_0=5 A_m$, and the blue solid lines are calculated under the ISA from Eq.~(\ref{eq:B}). }
\end{figure}

When an external magnetic field is applied during the DNP process, the evolution of the nuclear polarizations is changed drastically. For a single nuclear spin, it is easy to obtain the following analytical expression after $n$ DNP cycles, if the change of the nuclear polarization in a DNP cycle is small,
\begin{equation}
\label{eq:B}
p_n = 1-e^{-(A^2/\Omega^2)[1-\cos(\Omega\tau)]n/2},
\end{equation}
where $\Omega^2 = A^2 + \omega_0^2$. For a given $\tau$, the nuclear polarization $p_n$ decreases almost inverse-quadratically with oscillations as the Zeeman splitting of the external magnetic field $\omega_0$ increases. If the DNP cycle time $\tau$ is small enough, $\tau \ll \Omega^{-1}$, we find field-independent nuclear polarization
\begin{equation}
p_n \approx 1-e^{-A^2\tau^2n/4}.
\end{equation}

We present in Fig.~\ref{fig:B1} the analytical prediction from Eq.~(\ref{eq:B}) and numerical results, which agree well with each other for the given parameters. Such an  agreement between the analytical and numerical results indicates the validity of the adopted approximation, i.e., the change of the nuclear polarization in a DNP cycle is small. Compared to the zero field result, the nuclear polarization increases slower with an external magnetic field applied. Such a polarization suppression becomes severer in larger magnetic fields.

We shown in Fig.~\ref{fig:B4} the numerical results for $N=4$ nuclear spins and the analytical results from Eq.~(\ref{eq:B}) under the ISA. The good agreement between them indicates that the ISA is a good approximation. Compared with the zero magnetic field results, we observe the polarization suppression in a magnetic field. Furthermore, the suppression effect becomes more significant for the weakly coupled nuclear spins than that for the strongly coupled ones, due to the larger reduced magnetic field $\omega_0 / A_k$. In this sense, we may deduce that the application of a magnetic field during the DNP process actually enhance the inhomogeneity of the nuclear polarizations.

\section{Effect of nuclear dipolar interaction}
\label{sec:dnpD}

\begin{figure}
\includegraphics[width=3.25in]{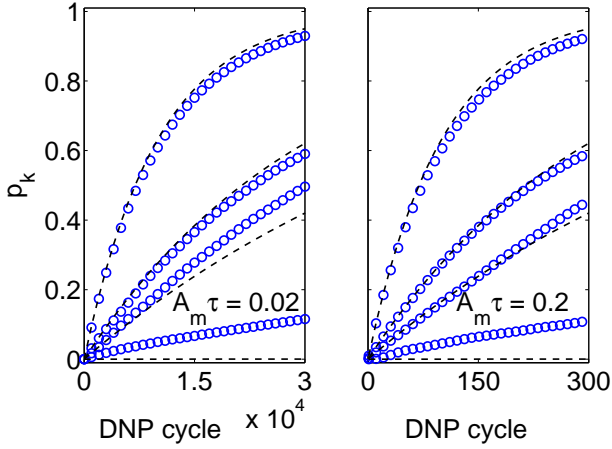}
\caption{\label{fig:dip4} (Color online) Effect of nuclear dipolar interaction on the nuclear polarization during DNP processes. The cycle times are $A_m \tau = 0.02$ (left) and $A_m \tau = 0.2$ (right). The dashed lines are the analytical results under ISA and the blue circles are numerical results. For those weakly coupled nuclear spins, spin diffusion induced by the dipolar coupling overwhelms the DNP induced by the hyperfine coupling.}
\end{figure}

The dipolar coupling between nuclear spins usually cause spin diffusion, i.e., the polarization are transferred from high polarization spins to low polarization spins. Since the nuclear polarization due to electron-nuclear hyperfine coupling during DNP process is highly nonuniform, including the nuclear dipolar coupling introduces a competing process which reverses or at least hinders the nuclear polarization inhomogeneity.

We compute the DNP process of $N=4$ nuclear spins, including both the hyperfine coupling and the nuclear dipolar coupling. The parameters are generated randomly: $A_1 = 0.7059$, $A_2=0.3009$, $A_3=0.0089$, $A_4=0.4014$, $\Gamma_{12} = 5.673\times 10^{-3}$, $\Gamma_{13} = 9.729\times 10^{-2}$, $\Gamma_{14}= 1.332\times 10^{-2}$, $\Gamma_{23}=6.490\times 10^{-2}$, $\Gamma_{24}=1.734\times 10^{-2}$, $\Gamma_{34}=3.909\times 10^{-2}$. The typical dipolar coupling strength is about one order of magnitude smaller than the typical hyperfine coupling strength.

We present the numerical results in Fig.~\ref{fig:dip4}, which shows clearly that the spin diffusion indeed occurs during the DNP process. Contrary to the usual free spin diffusion process, where the high polarizations decrease and the low polarizations increase with total nuclear polarization conserved, the spin diffusion during DNP does not cause the high polarization decrease while the low polarizations indeed increase with total nuclear polarization increasing. In other words, the dipolar interaction induced spin diffusion is more important for those weakly hyperfine-coupled nuclear spins whose polarization are mainly transferred from highly-polarized nuclear spins instead of the electron spin. Moreover, similarity in the left and right panel of Fig.~\ref{fig:dip4} indicates that reducing the DNP cycle period $\tau$ does not change significantly the final polarization distribution (inhomogeneity), except the spin diffusion process is slowed (the total evolution time $t = n \tau$ in the left panel of Fig.~\ref{fig:dip4} is 10 times as that in the right panel).

\section{Suppression of electron polarization decay via DNP effect}
\label{sec:dnpS}

\begin{figure}
\includegraphics[width=3.25in]{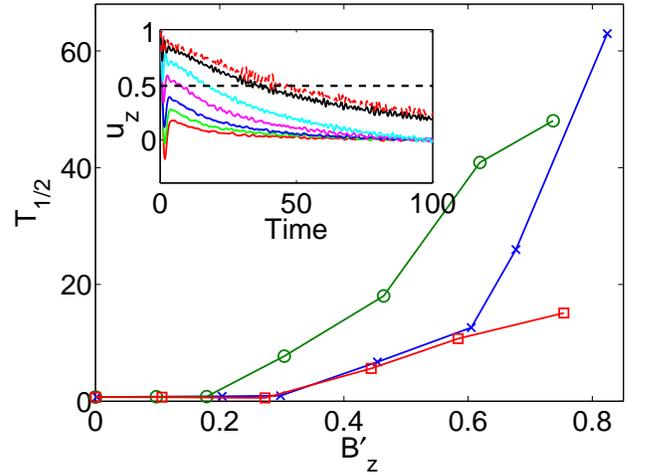}
\caption{\label{fig:tw} (Color online) Electron polarization decay time versus reduced Overhauser field for $N=256$ nuclear spins with Gaussian widths $\sigma_\perp = 0.2 N_\perp a$ (blue line with crosses), $0.4N_\perp a$ (green line with circles), and $0.6N_\perp a$ (red line with squares). The electron coherence time can be extended significantly even at moderate inhomogeneous nuclear polarizations (about $B_z' = 0.3$). Inset: Typical electron polarization decay for $\sigma_\perp = 0.4N_\perp a$ at different $\beta = 0, 5, 10, 20, 40, 80, 160$, from bottom to top. The dashed horizontal line denotes $u_z = 0.5$.}
\end{figure}

\begin{figure}
\includegraphics[width=3.25in]{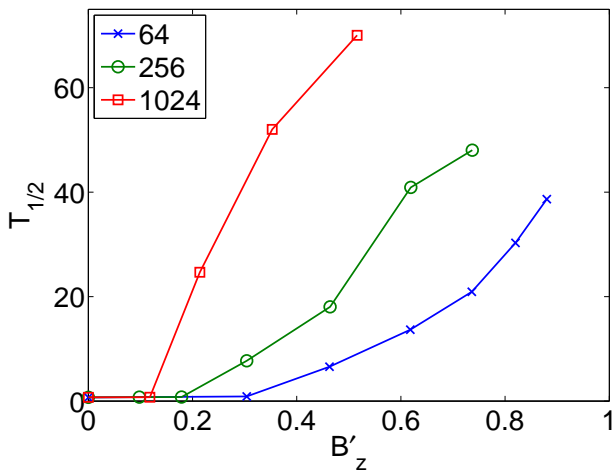}
\caption{\label{fig:tn} (Color online) The nuclear bath size effect on the electron polarization decay time. The electron density relative widths are the same for all three cases, $\sigma_\perp / (N_\perp a) = 0.4$. To reach the same decay time $T_{1/2}$, the required polarization of a large bath is smaller than that of a small bath.}
\end{figure}

DNP was originally proposed to extend the coherence time of the electron spin in the QD by suppressing the fluctuation of the nuclear-spin Overhauser field. The extension of the coherence time of electron spins in QDs was already demonstrated in double QD experiments and in some theoretical works~\cite{Reilly08, Ramon07, Ribeiro09, Stopa10, Bluhm10, Zhang10b, Gullans10, Gullans13}. However, as we show in Fig.~\ref{fig:12BDB} and~\ref{fig:BDB}, the fluctuation of the Overhauser field is not suppressed significantly after a DNP process, even at pretty high nuclear polarization. We also notice that in the experiment done by Bluhm {\it et al}.~\cite{Bluhm10} the suppression of the fluctuation of the Overhauser field is not inversely proportional to the extension of the coherence time, i.e., two thirds suppression of the fluctuation causes approximately seven times extension of $T_2^*$ (see Fig. 3 in Ref.~\cite{Bluhm10}). Therefore, other mechanisms, such as the protection effect of the nuclear spins strongly coupled to the electron and highly polarized during DNP~\cite{Zhang10b}, or the narrowing of the nuclear spin states~\cite{Klauser08, Bluhm10, Issler10, Asshoff11, Sun12}, might also play an important role.

We investigate the above conjecture by numerically simulating the electron spin polarization decay with an initial inhomogeneous nuclear polarization acquired during a previous DNP process. The coupling constants $A_k$ is assumed in a Gaussian distribution form
\begin{equation}
A_k = A_0 \exp\left(-\frac{(x_k-x_0)^2+(y_k-y_0)^2}{2\sigma_\perp^2}\right),
\end{equation}
where $A_0$ is the maximal coupling constant at the shifted center $(x_0,y_0) = (0.1, 0.27) a$, $(x_k, y_k) = (n_x-N_\perp, n_y-N_\perp) a$ is the position of the $k$th nuclear spin in a two-dimensional square lattice with $a$ being the lattice constant and $n_{x,y}\in [1,2N_\perp]$ the index of the $k$th nuclear spin. By increasing the width $\sigma_\perp$, we can adjust the $A_k$'s from a sharp distribution to a flat one.

The initial state is a product state of a fully polarized electron spin state and a partially and inhomogeneously polarized nuclear spin state. The electron spin is initially polarized along the $z$ direction. The nuclear spins are prepared through the DNP process as described in Sec.~\ref{sec:dnp}. We adopt the previously obtained ISA expression $p_k = 1-\exp(-2\beta A_k^2)$ with $\beta$ an adjustable parameter. For $\beta A_k^2\ll 1$, $p_k \approx 2\beta A_k^2$, which manifests the fact that $p_k$ is proportional to the square of the coupling constant. For $\beta A_k^2 \gg 1$, $p_k$ saturates at 1, which agrees with the limiting cases of an infinite number of DNP cycles.

The coupled electron-nuclear spin system evolves under the Hamiltonian described by Eq.~(\ref{eq:h}), but excluding the nuclear dipolar coupling and setting zero the Zeeman splitting of the external magnetic field. The numerical integration is performed with the P-representation density matrix method~\cite{Al-Hassanieh06, Zhang07r, Zhang10b}. This method has been numerically demonstrated as an efficient method for many-spin dynamics~\cite{Al-Hassanieh06, Zhang07r} and shows good agreement with the exact method based on Chebyshev polynomial expansion of the evolution operator, in describing the electron polarization decay~\cite{Zhang10b}.

The electron polarization $\langle 2S_z \rangle$ is monitored during the evolution. With a polarized nuclear spin bath which generates a nonzero Overhauser field in average, the electron polarization does not decay all the way down to zero. We thus subtract the nonvanishing value at very long time and normalize the decayed electron polarization as follows~\cite{Zhang10b}
\begin{equation}
\label{eq:ep}
u_z(t) = \frac{\langle S_z\rangle (t) - \langle S_z\rangle (\infty)}{\langle S_z\rangle (0)-\langle S_z\rangle (\infty)}
\end{equation}
where $\langle S_z\rangle (t) = \Tr\{S_z \rho(t)\}$ with $\rho(t)$ the density matrix of the coupled system at time $t$.

Typical electron polarization decay processes are shown in the inset of Fig.~\ref{fig:tw}. We see an initial quick dip at short times and slow decay at long times. To characterize the decay of $u_z$, we define a half-decay time constant $T_{1/2}$ at which $u_z = 1/2$. The relation between $T_{1/2}$ and the reduced Overhauser field $B'_z$ is presented in the main panel of  Fig.~\ref{fig:tw} for various distribution widths of $A_k$. At small $B'_z$, the half-decay time constant is almost independent of $B'_z$; while in a moderate region of $B'_z \sim 0.3$, the half-decay time increases rapidly and the growth rate becomes lager for a narrower distribution of $A_k$. This shows a big contrast to the relation between the field fluctuation $\Delta B'_z$ and $B'_z$ in Fig.~\ref{fig:12BDB} and~\ref{fig:BDB}, where $\Delta B'_z$ changes slowly while $B'_z$ increases.

To investigate the bath size effect on the electron polarization decay, we present in Fig.~\ref{fig:tn} the relation between $T_{1/2}$ and $B'_z$ for $N=64$, $256$, and $1024$. The results clearly show that the half-decay time $T_{1/2}$ increases more rapidly at moderate $B'_z$ for a larger bath size. More importantly, the transition position of $B'_z$, where the growth rate of $T_{1/2}$ changes abruptly, becomes smaller with the bath size increasing. Given this trend, the required $B'_z$ is far below 0.1 for a real QD with $N\sim 10^6$ nuclear spins if the electron polarization decay time is extended significantly.

To understand the mechanism of the electron polarization decay, we count the number of nuclear spins whose polarization is above a certain value $p_c$ in the case of $N=1024$ and $B'_z=0.52$ (the total nuclear polarization is about $22\%$). By setting $p_c = 95\%$, the number is $102$; by setting $p_c=90\%$, the number is $121$. Note also $T_{1/2} = 0.7$ at $B'_z = 0$ with the unpolarized initial nuclear state and $T_{1/2} = 70$ at $B'_z = 0.52$. The electron polarization decay time is extended about $100$ times, which roughly equals to the number of highly polarized (and strongly coupled with the electron) nuclear spins, indicating that these special spins protect the electron spin from decay~\cite{Zhang10b}.

\begin{figure}
\includegraphics[width=3.25in]{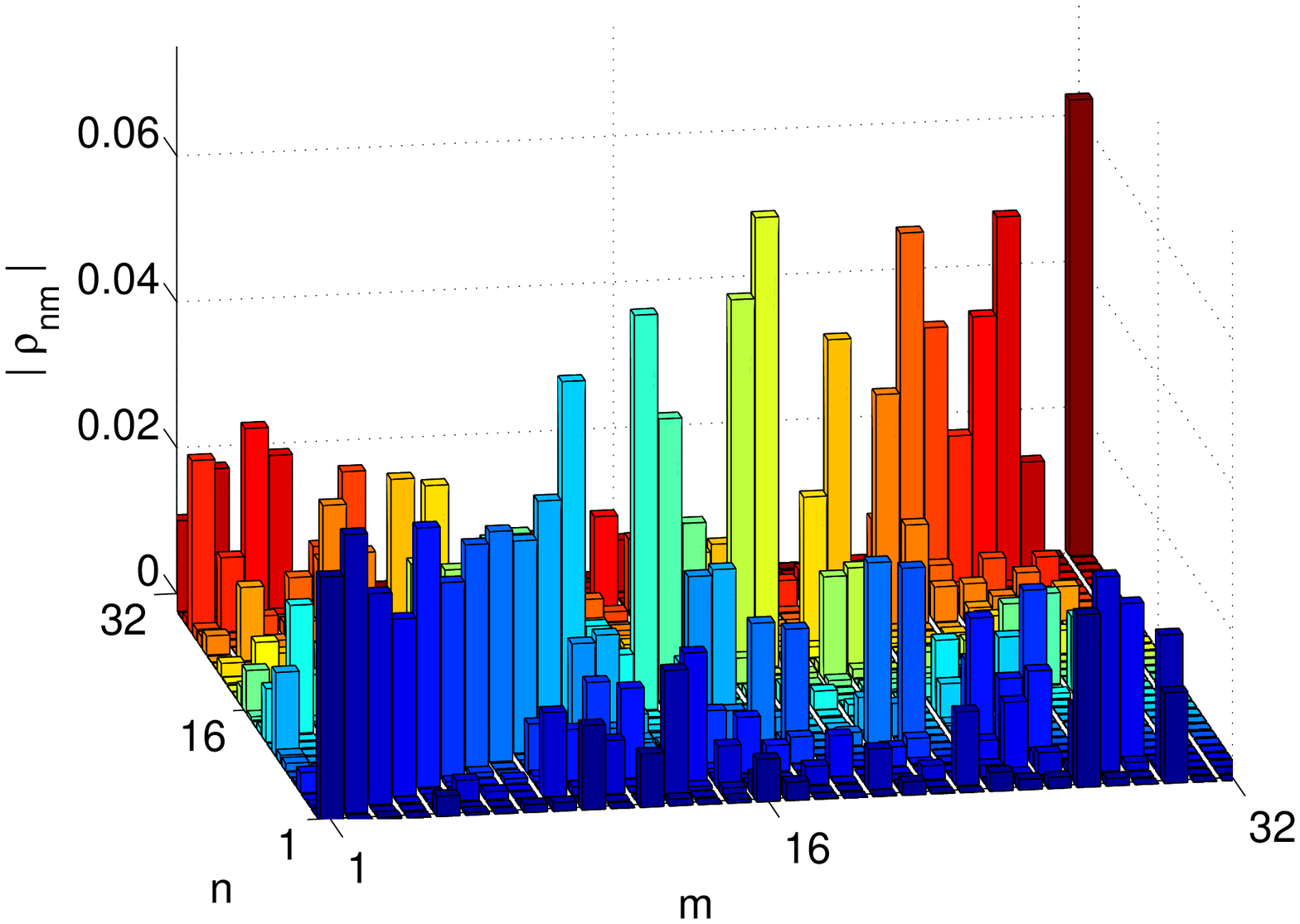}
\includegraphics[width=3.25in]{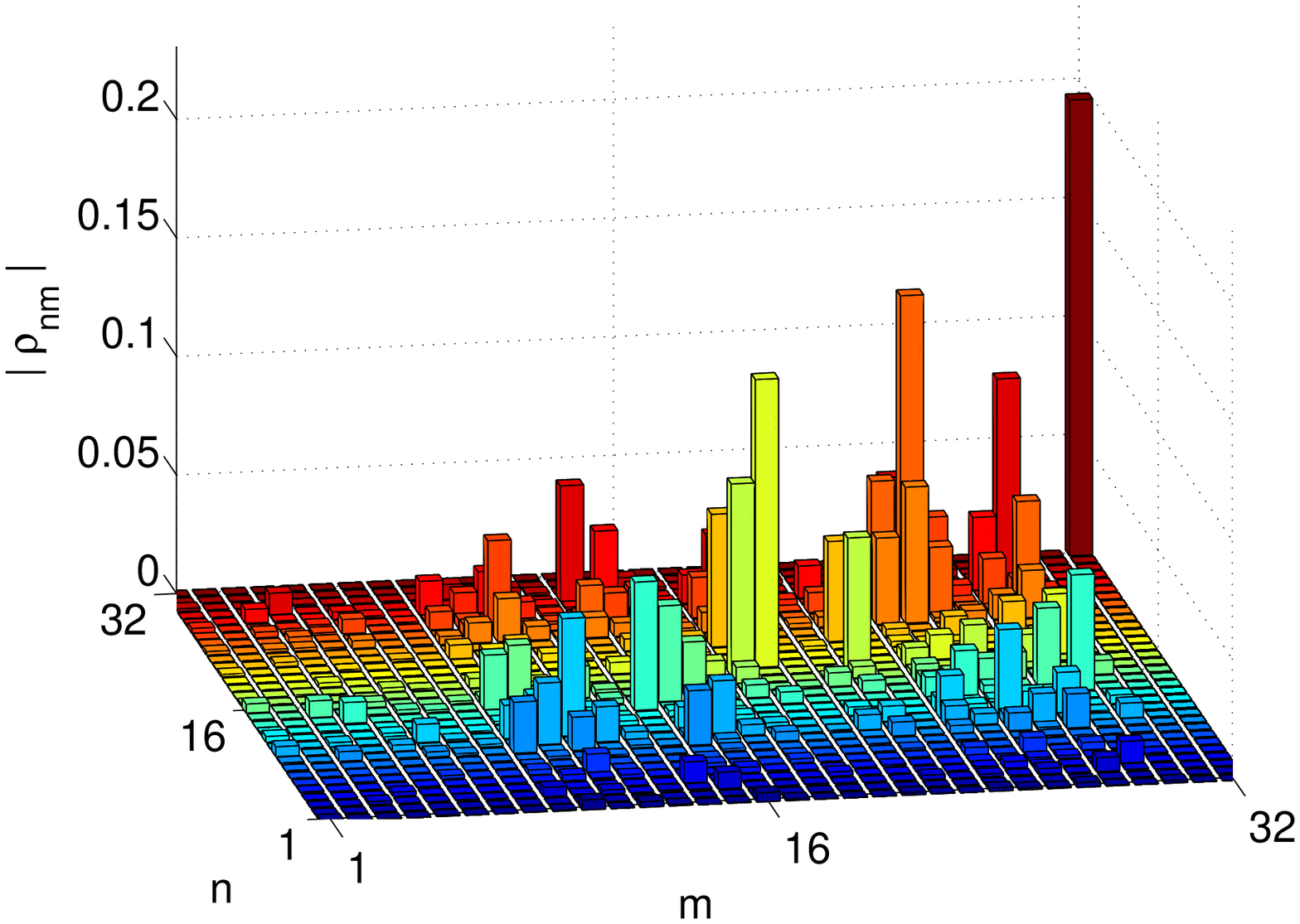}
\caption{\label{fig:rd} (Color online) Amplitudes of the Density matrix elements of the coupled electron-nuclear spin system in coupling basis before (top) and after (bottom) the DNP process. The number of nuclear spins is $N=4$. The density matrix is concentrated along the diagonal line, indicating a narrowing of the state distribution, through the DNP process.}
\end{figure}

Another mechanism to explain the extension of the electron polarization decay time is the nuclear state narrowing~\cite{Stepanenko06}. We plot in Fig.~\ref{fig:rd} the amplitude distribution of the density matrix elements before (top) and after (bottom) a DNP process for $N=4$ nuclear spins in its coupling basis, which is the eigen basis of the coupling Hamiltonian $H = {\bf S} \cdot \sum_k A_k {\bf I}_k$. Given a diagonal density matrix, the coupled electron-nuclear spin system would never change so that the decay time of the electron polarization is infinity. In this sense, the degree of the concentration of the density matrix on the diagonal line characterizes the effect of the narrowing state. By comparing the top and the bottom panel of Fig.~\ref{fig:rd}, we clearly see a narrowing of the state distribution after a DNP process, indicating qualitatively the extension of the decay time. But more detailed work need to be done in the future to interpret quantitatively the extension of the electron polarization decay.

\section{Conclusion}
\label{sec:con}

In summary, we obtain an analytical expression for the inhomogeneous nuclear polarization acquired during a DNP process under the independent nuclear spin approximation in a singly charged QD, where the nuclear spins coupled with an electron spin through contact hyperfine interaction. Numerical simulations agree well with the analytical predictions in the limit of short DNP cycles, due to the suppression of the electron-mediated nuclear spin interaction. The application of an external magnetic field does not change the overall DNP behaviors, except slowing the DNP process in proportion for all nuclear spins. The inclusion of the nuclear dipolar interaction causes weakly coupled nuclear spins slowly polarized through nuclear spin diffusion.

The inhomogeneously polarized nuclear spins after a DNP process are further utilized to extend the decay time of the electron polarization. Due to the protection effect of the highly polarized part of the nuclear spins, the electron polarization decay time can be extended 100 times at a rather low nuclear polarization (about 20\% for 1,000 nuclear spins and smaller for more nuclear spins).

\section{Acknowledgement}

This work was supported by he National Basic Research Program of China (Grant No. 2013CB922003), the National Natural Science Foundation of China under Grant No. 11275139 and 10904017, the Fundamental Research Funds for the Central Universities, and the Shanghai Pujiang Program under Grant No. 10PJ1401300.

\appendix
\section{Derivation of Eq.~(\ref{eq:ismb})}
\label{sec:appd}

From the definition of $B_z$, we obtain
\begin{eqnarray*}
B_z &=& \langle \sum_k A_k \hat I_{kz} \rangle \\
    &=& \sum_k A_k \langle \hat I_{kz} \rangle \\
    &=& {1\over 2} \sum_k A_k p_k.
\end{eqnarray*}

For the field fluctuation, we first calculate that
\begin{eqnarray*}
\langle \hat B_z^2 \rangle &=& \left\langle \sum_k \sum_j A_k A_j \hat I_{kz} \hat I_{jz}\right\rangle \\
    &=& \sum_k \sum_j A_k A_j \langle \hat I_{kz} \hat I_{jz}\rangle \\
    &=& {1\over 4} \left(\sum_k A_k^2 + \sum_k \sum_{j\neq k} A_k A_j p_k p_j \right).
\end{eqnarray*}
Then the field fluctuation becomes
\begin{eqnarray*}
\Delta B_z &=& \sqrt{\langle \hat B_z^2 \rangle - B_z^2} \\
    &=& \sqrt{{1\over 4} \sum_k (A_k^2 - A_k^2 p_k^2) }\\
    &=& {1\over 2} \sqrt{\sum_k A_k^2 (1- p_k^2)} \;.
\end{eqnarray*}



\begin{thebibliography}{55}
\expandafter\ifx\csname natexlab\endcsname\relax\def\natexlab#1{#1}\fi
\expandafter\ifx\csname bibnamefont\endcsname\relax
  \def\bibnamefont#1{#1}\fi
\expandafter\ifx\csname bibfnamefont\endcsname\relax
  \def\bibfnamefont#1{#1}\fi
\expandafter\ifx\csname citenamefont\endcsname\relax
  \def\citenamefont#1{#1}\fi
\expandafter\ifx\csname url\endcsname\relax
  \def\url#1{\texttt{#1}}\fi
\expandafter\ifx\csname urlprefix\endcsname\relax\def\urlprefix{URL }\fi
\providecommand{\bibinfo}[2]{#2}
\providecommand{\eprint}[2][]{\url{#2}}

\bibitem[{\citenamefont{Nielsen and Chuang}(2000)}]{Nielsen00}
\bibinfo{author}{\bibfnamefont{M.~A.} \bibnamefont{Nielsen}} \bibnamefont{and}
  \bibinfo{author}{\bibfnamefont{I.~L.} \bibnamefont{Chuang}},
  \emph{\bibinfo{title}{Quantum Computations and Quantum Information}}
  (\bibinfo{publisher}{Cambridge University Press, Cambridge},
  \bibinfo{year}{2000}).

\bibitem[{\citenamefont{Ladd et~al.}(2010)\citenamefont{Ladd, Jelezko,
  Laflamme, Nakamura, Monroe, and O'Brien}}]{Ladd10}
\bibinfo{author}{\bibfnamefont{T.~D.} \bibnamefont{Ladd}},
  \bibinfo{author}{\bibfnamefont{F.}~\bibnamefont{Jelezko}},
  \bibinfo{author}{\bibfnamefont{R.}~\bibnamefont{Laflamme}},
  \bibinfo{author}{\bibfnamefont{Y.}~\bibnamefont{Nakamura}},
  \bibinfo{author}{\bibfnamefont{C.}~\bibnamefont{Monroe}}, \bibnamefont{and}
  \bibinfo{author}{\bibfnamefont{J.~L.} \bibnamefont{O'Brien}},
  \bibinfo{journal}{Nature (London)} \textbf{\bibinfo{volume}{464}},
  \bibinfo{pages}{45} (\bibinfo{year}{2010}).

\bibitem[{\citenamefont{Zutic et~al.}(2004)\citenamefont{Zutic, Fabian, and
  Das~Sarma}}]{Zutic04}
\bibinfo{author}{\bibfnamefont{I.}~\bibnamefont{Zutic}},
  \bibinfo{author}{\bibfnamefont{J.}~\bibnamefont{Fabian}}, \bibnamefont{and}
  \bibinfo{author}{\bibfnamefont{S.}~\bibnamefont{Das~Sarma}},
  \bibinfo{journal}{Rev. Mod. Phys.} \textbf{\bibinfo{volume}{76}},
  \bibinfo{pages}{323} (\bibinfo{year}{2004}).

\bibitem[{\citenamefont{Slichter}(1992)}]{Slichter92}
\bibinfo{author}{\bibfnamefont{C.~P.} \bibnamefont{Slichter}},
  \emph{\bibinfo{title}{Principles of Magnetic Resonance}}
  (\bibinfo{publisher}{Springer-Verlag, New York}, \bibinfo{year}{1992}).

\bibitem[{\citenamefont{Viola et~al.}(1999)\citenamefont{Viola, Knill, and
  Lloyd}}]{Viola99}
\bibinfo{author}{\bibfnamefont{L.}~\bibnamefont{Viola}},
  \bibinfo{author}{\bibfnamefont{E.}~\bibnamefont{Knill}}, \bibnamefont{and}
  \bibinfo{author}{\bibfnamefont{S.}~\bibnamefont{Lloyd}},
  \bibinfo{journal}{Phys. Rev. Lett.} \textbf{\bibinfo{volume}{82}},
  \bibinfo{pages}{2417} (\bibinfo{year}{1999}).

\bibitem[{\citenamefont{Viola et~al.}(2000)\citenamefont{Viola, Knill, and
  Lloyd}}]{Viola00}
\bibinfo{author}{\bibfnamefont{L.}~\bibnamefont{Viola}},
  \bibinfo{author}{\bibfnamefont{E.}~\bibnamefont{Knill}}, \bibnamefont{and}
  \bibinfo{author}{\bibfnamefont{S.}~\bibnamefont{Lloyd}},
  \bibinfo{journal}{Phys. Rev. Lett.} \textbf{\bibinfo{volume}{85}},
  \bibinfo{pages}{3520} (\bibinfo{year}{2000}).

\bibitem[{\citenamefont{Khodjasteh and Lidar}(2005)}]{Khodjasteh05}
\bibinfo{author}{\bibfnamefont{K.}~\bibnamefont{Khodjasteh}} \bibnamefont{and}
  \bibinfo{author}{\bibfnamefont{D.~A.} \bibnamefont{Lidar}},
  \bibinfo{journal}{Phys. Rev. Lett.} \textbf{\bibinfo{volume}{95}},
  \bibinfo{eid}{180501} (\bibinfo{year}{2005}).

\bibitem[{\citenamefont{Scully and Zubairy}(1997)}]{Scully97}
\bibinfo{author}{\bibfnamefont{M.~O.} \bibnamefont{Scully}} \bibnamefont{and}
  \bibinfo{author}{\bibfnamefont{M.~S.} \bibnamefont{Zubairy}},
  \emph{\bibinfo{title}{Quantum Optics}} (\bibinfo{publisher}{Cambridge
  University Press, Cambridge}, \bibinfo{year}{1997}).

\bibitem[{\citenamefont{Stepanenko et~al.}(2006)\citenamefont{Stepanenko,
  Burkard, Giedke, and Imamoglu}}]{Stepanenko06}
\bibinfo{author}{\bibfnamefont{D.}~\bibnamefont{Stepanenko}},
  \bibinfo{author}{\bibfnamefont{G.}~\bibnamefont{Burkard}},
  \bibinfo{author}{\bibfnamefont{G.}~\bibnamefont{Giedke}}, \bibnamefont{and}
  \bibinfo{author}{\bibfnamefont{A.}~\bibnamefont{Imamoglu}},
  \bibinfo{journal}{Phys. Rev. Lett.} \textbf{\bibinfo{volume}{96}},
  \bibinfo{eid}{136401} (\bibinfo{year}{2006}).

\bibitem[{\citenamefont{Ribeiro and Burkard}(2009)}]{Ribeiro09}
\bibinfo{author}{\bibfnamefont{H.}~\bibnamefont{Ribeiro}} \bibnamefont{and}
  \bibinfo{author}{\bibfnamefont{G.}~\bibnamefont{Burkard}},
  \bibinfo{journal}{Phys. Rev. Lett.} \textbf{\bibinfo{volume}{102}},
  \bibinfo{eid}{216802} (\bibinfo{year}{2009}).

\bibitem[{\citenamefont{Yao}(2011)}]{Yao11}
\bibinfo{author}{\bibfnamefont{W.}~\bibnamefont{Yao}}, \bibinfo{journal}{Phys.
  Rev. B} \textbf{\bibinfo{volume}{83}}, \bibinfo{pages}{201308(R)}
  (\bibinfo{year}{2011}).

\bibitem[{\citenamefont{Loss and DiVincenzo}(1998)}]{Loss98}
\bibinfo{author}{\bibfnamefont{D.}~\bibnamefont{Loss}} \bibnamefont{and}
  \bibinfo{author}{\bibfnamefont{D.~P.} \bibnamefont{DiVincenzo}},
  \bibinfo{journal}{Phys. Rev. A} \textbf{\bibinfo{volume}{57}},
  \bibinfo{pages}{120} (\bibinfo{year}{1998}).

\bibitem[{\citenamefont{Kane}(1998)}]{Kane98}
\bibinfo{author}{\bibfnamefont{B.}~\bibnamefont{Kane}},
  \bibinfo{journal}{Nature (London)} \textbf{\bibinfo{volume}{393}},
  \bibinfo{pages}{133} (\bibinfo{year}{1998}).

\bibitem[{\citenamefont{Petta et~al.}(2005)\citenamefont{Petta, Johnson,
  Taylor, Laird, Yacoby, Lukin, Marcus, Hanson, and Gossard}}]{Petta05}
\bibinfo{author}{\bibfnamefont{J.~R.} \bibnamefont{Petta}},
  \bibinfo{author}{\bibfnamefont{A.~C.} \bibnamefont{Johnson}},
  \bibinfo{author}{\bibfnamefont{J.}~\bibnamefont{Taylor}},
  \bibinfo{author}{\bibfnamefont{E.~A.} \bibnamefont{Laird}},
  \bibinfo{author}{\bibfnamefont{A.}~\bibnamefont{Yacoby}},
  \bibinfo{author}{\bibfnamefont{M.~D.} \bibnamefont{Lukin}},
  \bibinfo{author}{\bibfnamefont{C.~M.} \bibnamefont{Marcus}},
  \bibinfo{author}{\bibfnamefont{M.~P.} \bibnamefont{Hanson}},
  \bibnamefont{and} \bibinfo{author}{\bibfnamefont{A.~C.}
  \bibnamefont{Gossard}}, \bibinfo{journal}{Science}
  \textbf{\bibinfo{volume}{309}}, \bibinfo{pages}{2180} (\bibinfo{year}{2005}).

\bibitem[{\citenamefont{Koppens et~al.}(2005)\citenamefont{Koppens, Folk,
  Elzerman, Hanson, van Beveren, Vink, Tranitz, Wegscheider, Kouwenhoven, and
  Vandersypen}}]{Koppens05}
\bibinfo{author}{\bibfnamefont{F.~H.~L.} \bibnamefont{Koppens}},
  \bibinfo{author}{\bibfnamefont{J.~A.} \bibnamefont{Folk}},
  \bibinfo{author}{\bibfnamefont{J.~M.} \bibnamefont{Elzerman}},
  \bibinfo{author}{\bibfnamefont{R.}~\bibnamefont{Hanson}},
  \bibinfo{author}{\bibfnamefont{L.~H.~W.} \bibnamefont{van Beveren}},
  \bibinfo{author}{\bibfnamefont{I.~T.} \bibnamefont{Vink}},
  \bibinfo{author}{\bibfnamefont{H.~P.} \bibnamefont{Tranitz}},
  \bibinfo{author}{\bibfnamefont{W.}~\bibnamefont{Wegscheider}},
  \bibinfo{author}{\bibfnamefont{L.~P.} \bibnamefont{Kouwenhoven}},
  \bibnamefont{and} \bibinfo{author}{\bibfnamefont{L.~M.~K.}
  \bibnamefont{Vandersypen}}, \bibinfo{journal}{Science}
  \textbf{\bibinfo{volume}{309}}, \bibinfo{pages}{1346} (\bibinfo{year}{2005}).

\bibitem[{\citenamefont{Johnson et~al.}(2005)\citenamefont{Johnson, Petta,
  Taylor, Yacoby, Lukin, Marcus, Hanson, and Gossard}}]{Johnson05}
\bibinfo{author}{\bibfnamefont{A.~C.} \bibnamefont{Johnson}},
  \bibinfo{author}{\bibfnamefont{J.~R.} \bibnamefont{Petta}},
  \bibinfo{author}{\bibfnamefont{J.~M.} \bibnamefont{Taylor}},
  \bibinfo{author}{\bibfnamefont{A.}~\bibnamefont{Yacoby}},
  \bibinfo{author}{\bibfnamefont{M.~D.} \bibnamefont{Lukin}},
  \bibinfo{author}{\bibfnamefont{C.~M.} \bibnamefont{Marcus}},
  \bibinfo{author}{\bibfnamefont{M.~P.} \bibnamefont{Hanson}},
  \bibnamefont{and} \bibinfo{author}{\bibfnamefont{A.~C.}
  \bibnamefont{Gossard}}, \bibinfo{journal}{Nature (London)}
  \textbf{\bibinfo{volume}{435}}, \bibinfo{pages}{925} (\bibinfo{year}{2005}).

\bibitem[{\citenamefont{Merkulov et~al.}(2002)\citenamefont{Merkulov, Efros,
  and Rosen}}]{Merkulov02}
\bibinfo{author}{\bibfnamefont{I.~A.} \bibnamefont{Merkulov}},
  \bibinfo{author}{\bibfnamefont{A.~L.} \bibnamefont{Efros}}, \bibnamefont{and}
  \bibinfo{author}{\bibfnamefont{M.}~\bibnamefont{Rosen}},
  \bibinfo{journal}{Phys. Rev. B} \textbf{\bibinfo{volume}{65}},
  \bibinfo{pages}{205309} (\bibinfo{year}{2002}).

\bibitem[{\citenamefont{Zhang et~al.}(2006)\citenamefont{Zhang, Dobrovitski,
  Al-Hassanieh, Dagotto, and Harmon}}]{Zhang06}
\bibinfo{author}{\bibfnamefont{W.}~\bibnamefont{Zhang}},
  \bibinfo{author}{\bibfnamefont{V.~V.} \bibnamefont{Dobrovitski}},
  \bibinfo{author}{\bibfnamefont{K.~A.} \bibnamefont{Al-Hassanieh}},
  \bibinfo{author}{\bibfnamefont{E.}~\bibnamefont{Dagotto}}, \bibnamefont{and}
  \bibinfo{author}{\bibfnamefont{B.~N.} \bibnamefont{Harmon}},
  \bibinfo{journal}{Phys. Rev. B} \textbf{\bibinfo{volume}{74}},
  \bibinfo{pages}{205313} (\bibinfo{year}{2006}).

\bibitem[{\citenamefont{Deng and Hu}(2006)}]{Deng06}
\bibinfo{author}{\bibfnamefont{C.}~\bibnamefont{Deng}} \bibnamefont{and}
  \bibinfo{author}{\bibfnamefont{X.}~\bibnamefont{Hu}}, \bibinfo{journal}{Phys.
  Rev. B} \textbf{\bibinfo{volume}{73}}, \bibinfo{pages}{241303(R)}
  (\bibinfo{year}{2006}).

\bibitem[{\citenamefont{Greilich et~al.}(2007)\citenamefont{Greilich, Shabaev,
  Yakovlev, Efros, Yugova, Reuter, Wieck, and Bayer}}]{Greilich07}
\bibinfo{author}{\bibfnamefont{A.}~\bibnamefont{Greilich}},
  \bibinfo{author}{\bibfnamefont{A.}~\bibnamefont{Shabaev}},
  \bibinfo{author}{\bibfnamefont{D.~R.} \bibnamefont{Yakovlev}},
  \bibinfo{author}{\bibfnamefont{A.~L.} \bibnamefont{Efros}},
  \bibinfo{author}{\bibfnamefont{I.~A.} \bibnamefont{Yugova}},
  \bibinfo{author}{\bibfnamefont{D.}~\bibnamefont{Reuter}},
  \bibinfo{author}{\bibfnamefont{A.~D.} \bibnamefont{Wieck}}, \bibnamefont{and}
  \bibinfo{author}{\bibfnamefont{M.}~\bibnamefont{Bayer}},
  \bibinfo{journal}{Science} \textbf{\bibinfo{volume}{317}},
  \bibinfo{pages}{1896} (\bibinfo{year}{2007}).

\bibitem[{\citenamefont{Reilly et~al.}(2008)\citenamefont{Reilly, Taylor,
  Petta, Marcus, Hanson, and Gossard}}]{Reilly08}
\bibinfo{author}{\bibfnamefont{D.~J.} \bibnamefont{Reilly}},
  \bibinfo{author}{\bibfnamefont{J.~M.} \bibnamefont{Taylor}},
  \bibinfo{author}{\bibfnamefont{J.~R.} \bibnamefont{Petta}},
  \bibinfo{author}{\bibfnamefont{C.~M.} \bibnamefont{Marcus}},
  \bibinfo{author}{\bibfnamefont{M.~P.} \bibnamefont{Hanson}},
  \bibnamefont{and} \bibinfo{author}{\bibfnamefont{A.~C.}
  \bibnamefont{Gossard}}, \bibinfo{journal}{Science}
  \textbf{\bibinfo{volume}{321}}, \bibinfo{pages}{817} (\bibinfo{year}{2008}).

\bibitem[{\citenamefont{Xu et~al.}(2009)\citenamefont{Xu, Yao, Sun, Steel,
  Bracker, Gammon, and Sham}}]{Xu09}
\bibinfo{author}{\bibfnamefont{X.}~\bibnamefont{Xu}},
  \bibinfo{author}{\bibfnamefont{W.}~\bibnamefont{Yao}},
  \bibinfo{author}{\bibfnamefont{B.}~\bibnamefont{Sun}},
  \bibinfo{author}{\bibfnamefont{D.~G.} \bibnamefont{Steel}},
  \bibinfo{author}{\bibfnamefont{A.~S.} \bibnamefont{Bracker}},
  \bibinfo{author}{\bibfnamefont{D.}~\bibnamefont{Gammon}}, \bibnamefont{and}
  \bibinfo{author}{\bibfnamefont{L.~J.} \bibnamefont{Sham}},
  \bibinfo{journal}{Nature (London)} \textbf{\bibinfo{volume}{459}},
  \bibinfo{eid}{1105} (\bibinfo{year}{2009}).

\bibitem[{\citenamefont{Yao et~al.}(2007)\citenamefont{Yao, Liu, and
  Sham}}]{Yao07}
\bibinfo{author}{\bibfnamefont{W.}~\bibnamefont{Yao}},
  \bibinfo{author}{\bibfnamefont{R.-B.} \bibnamefont{Liu}}, \bibnamefont{and}
  \bibinfo{author}{\bibfnamefont{L.~J.} \bibnamefont{Sham}},
  \bibinfo{journal}{Phys. Rev. Lett.} \textbf{\bibinfo{volume}{98}},
  \bibinfo{eid}{077602} (\bibinfo{year}{2007}).

\bibitem[{\citenamefont{Witzel and Sarma}(2007)}]{Witzel07}
\bibinfo{author}{\bibfnamefont{W.~M.} \bibnamefont{Witzel}} \bibnamefont{and}
  \bibinfo{author}{\bibfnamefont{S.} \bibnamefont{Das Sarma}},
  \bibinfo{journal}{Phys. Rev. Lett.} \textbf{\bibinfo{volume}{98}},
  \bibinfo{eid}{077601} (\bibinfo{year}{2007}).

\bibitem[{\citenamefont{Zhang et~al.}(2007{\natexlab{a}})\citenamefont{Zhang,
  Dobrovitski, Santos, Viola, and Harmon}}]{Zhang07a}
\bibinfo{author}{\bibfnamefont{W.}~\bibnamefont{Zhang}},
  \bibinfo{author}{\bibfnamefont{V.~V.} \bibnamefont{Dobrovitski}},
  \bibinfo{author}{\bibfnamefont{L.~F.} \bibnamefont{Santos}},
  \bibinfo{author}{\bibfnamefont{L.}~\bibnamefont{Viola}}, \bibnamefont{and}
  \bibinfo{author}{\bibfnamefont{B.~N.} \bibnamefont{Harmon}},
  \bibinfo{journal}{Phys. Rev. B} \textbf{\bibinfo{volume}{75}},
  \bibinfo{pages}{201302(R)} (\bibinfo{year}{2007}{\natexlab{a}}).

\bibitem[{\citenamefont{Khodjasteh and Lidar}(2007)}]{Khodjasteh07}
\bibinfo{author}{\bibfnamefont{K.}~\bibnamefont{Khodjasteh}} \bibnamefont{and}
  \bibinfo{author}{\bibfnamefont{D.~A.} \bibnamefont{Lidar}},
  \bibinfo{journal}{Phys. Rev. A} \textbf{\bibinfo{volume}{75}},
  \bibinfo{pages}{062310} (\bibinfo{year}{2007}).

\bibitem[{\citenamefont{Zhang et~al.}(2008)\citenamefont{Zhang, Konstantinidis,
  Dobrovitski, Harmon, Santos, and Viola}}]{Zhang08}
\bibinfo{author}{\bibfnamefont{W.}~\bibnamefont{Zhang}},
  \bibinfo{author}{\bibfnamefont{N.~P.} \bibnamefont{Konstantinidis}},
  \bibinfo{author}{\bibfnamefont{V.~V.} \bibnamefont{Dobrovitski}},
  \bibinfo{author}{\bibfnamefont{B.~N.} \bibnamefont{Harmon}},
  \bibinfo{author}{\bibfnamefont{L.~F.} \bibnamefont{Santos}},
  \bibnamefont{and} \bibinfo{author}{\bibfnamefont{L.}~\bibnamefont{Viola}},
  \bibinfo{journal}{Phys. Rev. B} \textbf{\bibinfo{volume}{77}},
  \bibinfo{pages}{125336} (\bibinfo{year}{2008}).

\bibitem[{\citenamefont{Klauser et~al.}(2006)\citenamefont{Klauser, Coish, and
  Loss}}]{Klauser06}
\bibinfo{author}{\bibfnamefont{D.}~\bibnamefont{Klauser}},
  \bibinfo{author}{\bibfnamefont{W.~A.} \bibnamefont{Coish}}, \bibnamefont{and}
  \bibinfo{author}{\bibfnamefont{D.}~\bibnamefont{Loss}},
  \bibinfo{journal}{Phys. Rev. B} \textbf{\bibinfo{volume}{73}},
  \bibinfo{eid}{205302} (\bibinfo{year}{2006}).

\bibitem[{\citenamefont{Ramon and Hu}(2007)}]{Ramon07}
\bibinfo{author}{\bibfnamefont{G.}~\bibnamefont{Ramon}} \bibnamefont{and}
  \bibinfo{author}{\bibfnamefont{X.}~\bibnamefont{Hu}}, \bibinfo{journal}{Phys.
  Rev. B} \textbf{\bibinfo{volume}{75}}, \bibinfo{pages}{161301(R)}
  (\bibinfo{year}{2007}).

\bibitem[{\citenamefont{Zhang et~al.}(2007{\natexlab{b}})\citenamefont{Zhang,
  Dobrovitski, Santos, Viola, and Harmon}}]{Zhang07b}
\bibinfo{author}{\bibfnamefont{W.}~\bibnamefont{Zhang}},
  \bibinfo{author}{\bibfnamefont{V.~V.} \bibnamefont{Dobrovitski}},
  \bibinfo{author}{\bibfnamefont{L.~F.} \bibnamefont{Santos}},
  \bibinfo{author}{\bibfnamefont{L.}~\bibnamefont{Viola}}, \bibnamefont{and}
  \bibinfo{author}{\bibfnamefont{B.~N.} \bibnamefont{Harmon}},
  \bibinfo{journal}{J. Mod. Opt.} \textbf{\bibinfo{volume}{54}},
  \bibinfo{pages}{2629} (\bibinfo{year}{2007}{\natexlab{b}}).

\bibitem[{\citenamefont{Burkard et~al.}(1999)\citenamefont{Burkard, Loss, and
  DiVincenzo}}]{Burkard99}
\bibinfo{author}{\bibfnamefont{G.}~\bibnamefont{Burkard}},
  \bibinfo{author}{\bibfnamefont{D.}~\bibnamefont{Loss}}, \bibnamefont{and}
  \bibinfo{author}{\bibfnamefont{D.~P.} \bibnamefont{DiVincenzo}},
  \bibinfo{journal}{Phys. Rev. B} \textbf{\bibinfo{volume}{59}},
  \bibinfo{pages}{2070} (\bibinfo{year}{1999}).

\bibitem[{\citenamefont{Coish and Loss}(2004)}]{Coish04}
\bibinfo{author}{\bibfnamefont{W.~A.} \bibnamefont{Coish}} \bibnamefont{and}
  \bibinfo{author}{\bibfnamefont{D.}~\bibnamefont{Loss}},
  \bibinfo{journal}{Phys. Rev. B} \textbf{\bibinfo{volume}{70}},
  \bibinfo{eid}{195340} (\bibinfo{year}{2004}).

\bibitem[{\citenamefont{Brown et~al.}(1996)\citenamefont{Brown, Kennedy,
  Gammon, and Snow}}]{Brown96}
\bibinfo{author}{\bibfnamefont{S.~W.} \bibnamefont{Brown}},
  \bibinfo{author}{\bibfnamefont{T.~A.} \bibnamefont{Kennedy}},
  \bibinfo{author}{\bibfnamefont{D.}~\bibnamefont{Gammon}}, \bibnamefont{and}
  \bibinfo{author}{\bibfnamefont{E.~S.} \bibnamefont{Snow}},
  \bibinfo{journal}{Phys. Rev. B} \textbf{\bibinfo{volume}{54}},
  \bibinfo{pages}{R17339} (\bibinfo{year}{1996}).

\bibitem[{\citenamefont{Bracker et~al.}(2005)\citenamefont{Bracker, Stinaff,
  Gammon, Ware, Tischler, Shabaev, Efros, Park, Gershoni, Korenev
  et~al.}}]{Bracker05}
\bibinfo{author}{\bibfnamefont{A.}~\bibnamefont{Bracker}},
  \bibinfo{author}{\bibfnamefont{E.}~\bibnamefont{Stinaff}},
  \bibinfo{author}{\bibfnamefont{D.}~\bibnamefont{Gammon}},
  \bibinfo{author}{\bibfnamefont{M.}~\bibnamefont{Ware}},
  \bibinfo{author}{\bibfnamefont{J.}~\bibnamefont{Tischler}},
  \bibinfo{author}{\bibfnamefont{A.}~\bibnamefont{Shabaev}},
  \bibinfo{author}{\bibfnamefont{A.}~\bibnamefont{Efros}},
  \bibinfo{author}{\bibfnamefont{D.}~\bibnamefont{Park}},
  \bibinfo{author}{\bibfnamefont{D.}~\bibnamefont{Gershoni}},
  \bibinfo{author}{\bibfnamefont{V.}~\bibnamefont{Korenev}},
  \bibnamefont{et~al.}, \bibinfo{journal}{Phys. Rev. Lett.}
  \textbf{\bibinfo{volume}{94}}, \bibinfo{pages}{047402}
  (\bibinfo{year}{2005}).

\bibitem[{\citenamefont{Ono et~al.}(2002)\citenamefont{Ono, Austing, Tokura,
  and Tarucha}}]{Ono02}
\bibinfo{author}{\bibfnamefont{K.}~\bibnamefont{Ono}},
  \bibinfo{author}{\bibfnamefont{D.}~\bibnamefont{Austing}},
  \bibinfo{author}{\bibfnamefont{Y.}~\bibnamefont{Tokura}}, \bibnamefont{and}
  \bibinfo{author}{\bibfnamefont{S.}~\bibnamefont{Tarucha}},
  \bibinfo{journal}{Science} \textbf{\bibinfo{volume}{297}},
  \bibinfo{pages}{1313} (\bibinfo{year}{2002}).

\bibitem[{\citenamefont{Petta et~al.}(2008)\citenamefont{Petta, Taylor,
  Johnson, Yacoby, Lukin, Marcus, Hanson, and Gossard}}]{Petta08}
\bibinfo{author}{\bibfnamefont{J.~R.} \bibnamefont{Petta}},
  \bibinfo{author}{\bibfnamefont{J.~M.} \bibnamefont{Taylor}},
  \bibinfo{author}{\bibfnamefont{A.~C.} \bibnamefont{Johnson}},
  \bibinfo{author}{\bibfnamefont{A.}~\bibnamefont{Yacoby}},
  \bibinfo{author}{\bibfnamefont{M.~D.} \bibnamefont{Lukin}},
  \bibinfo{author}{\bibfnamefont{C.~M.} \bibnamefont{Marcus}},
  \bibinfo{author}{\bibfnamefont{M.~P.} \bibnamefont{Hanson}},
  \bibnamefont{and} \bibinfo{author}{\bibfnamefont{A.~C.}
  \bibnamefont{Gossard}}, \bibinfo{journal}{Phys. Rev. Lett.}
  \textbf{\bibinfo{volume}{100}}, \bibinfo{pages}{067601}
  (\bibinfo{year}{2008}).

\bibitem[{\citenamefont{Reilly et~al.}(2010)\citenamefont{Reilly, Taylor,
  Petta, Marcus, Hanson, and Gossard}}]{Reilly10}
\bibinfo{author}{\bibfnamefont{D.~J.} \bibnamefont{Reilly}},
  \bibinfo{author}{\bibfnamefont{J.~M.} \bibnamefont{Taylor}},
  \bibinfo{author}{\bibfnamefont{J.~R.} \bibnamefont{Petta}},
  \bibinfo{author}{\bibfnamefont{C.~M.} \bibnamefont{Marcus}},
  \bibinfo{author}{\bibfnamefont{M.~P.} \bibnamefont{Hanson}},
  \bibnamefont{and} \bibinfo{author}{\bibfnamefont{A.~C.}
  \bibnamefont{Gossard}}, \bibinfo{journal}{Phys. Rev. Lett.}
  \textbf{\bibinfo{volume}{104}}, \bibinfo{pages}{236802}
  (\bibinfo{year}{2010}).

\bibitem[{\citenamefont{Issler et~al.}(2010)\citenamefont{Issler, Kessler,
  Giedke, Yelin, Cirac, Lukin, and Imamoglu}}]{Issler10}
\bibinfo{author}{\bibfnamefont{M.}~\bibnamefont{Issler}},
  \bibinfo{author}{\bibfnamefont{E.~M.} \bibnamefont{Kessler}},
  \bibinfo{author}{\bibfnamefont{G.}~\bibnamefont{Giedke}},
  \bibinfo{author}{\bibfnamefont{S.}~\bibnamefont{Yelin}},
  \bibinfo{author}{\bibfnamefont{I.}~\bibnamefont{Cirac}},
  \bibinfo{author}{\bibfnamefont{M.~D.} \bibnamefont{Lukin}}, \bibnamefont{and}
  \bibinfo{author}{\bibfnamefont{A.}~\bibnamefont{Imamoglu}},
  \bibinfo{journal}{Phys. Rev. Lett.} \textbf{\bibinfo{volume}{105}},
  \bibinfo{pages}{267202} (\bibinfo{year}{2010}).

\bibitem[{\citenamefont{Sun et~al.}(2012)\citenamefont{Sun, Chow, Steel,
  Bracker, Gammon, and Sham}}]{Sun12}
\bibinfo{author}{\bibfnamefont{B.}~\bibnamefont{Sun}},
  \bibinfo{author}{\bibfnamefont{C.~M.~E.} \bibnamefont{Chow}},
  \bibinfo{author}{\bibfnamefont{D.~G.} \bibnamefont{Steel}},
  \bibinfo{author}{\bibfnamefont{A.~S.} \bibnamefont{Bracker}},
  \bibinfo{author}{\bibfnamefont{D.}~\bibnamefont{Gammon}}, \bibnamefont{and}
  \bibinfo{author}{\bibfnamefont{L.~J.} \bibnamefont{Sham}},
  \bibinfo{journal}{Phys. Rev. Lett.} \textbf{\bibinfo{volume}{108}},
  \bibinfo{pages}{187401} (\bibinfo{year}{2012}).

\bibitem[{\citenamefont{Yang and Sham}(2012)}]{Yang12}
\bibinfo{author}{\bibfnamefont{W.}~\bibnamefont{Yang}} \bibnamefont{and}
  \bibinfo{author}{\bibfnamefont{L.~J.} \bibnamefont{Sham}},
  \bibinfo{journal}{Phys. Rev. B} \textbf{\bibinfo{volume}{85}},
  \bibinfo{pages}{235319} (\bibinfo{year}{2012}).

\bibitem[{\citenamefont{Barthel et~al.}(2012)\citenamefont{Barthel, Medford,
  Bluhm, Yacoby, Marcus, Hanson, and Gossard}}]{Barthel12}
\bibinfo{author}{\bibfnamefont{C.}~\bibnamefont{Barthel}},
  \bibinfo{author}{\bibfnamefont{J.}~\bibnamefont{Medford}},
  \bibinfo{author}{\bibfnamefont{H.}~\bibnamefont{Bluhm}},
  \bibinfo{author}{\bibfnamefont{A.}~\bibnamefont{Yacoby}},
  \bibinfo{author}{\bibfnamefont{C.~M.} \bibnamefont{Marcus}},
  \bibinfo{author}{\bibfnamefont{M.~P.} \bibnamefont{Hanson}},
  \bibnamefont{and} \bibinfo{author}{\bibfnamefont{A.~C.}
  \bibnamefont{Gossard}}, \bibinfo{journal}{Phys. Rev. B}
  \textbf{\bibinfo{volume}{85}}, \bibinfo{pages}{035306}
  (\bibinfo{year}{2012}).

\bibitem[{\citenamefont{Foletti et~al.}(2009)\citenamefont{Foletti, Bluhm,
  Mahalu, Umansky, and Yacoby}}]{Foletti09}
\bibinfo{author}{\bibfnamefont{S.}~\bibnamefont{Foletti}},
  \bibinfo{author}{\bibfnamefont{H.}~\bibnamefont{Bluhm}},
  \bibinfo{author}{\bibfnamefont{D.}~\bibnamefont{Mahalu}},
  \bibinfo{author}{\bibfnamefont{V.}~\bibnamefont{Umansky}}, \bibnamefont{and}
  \bibinfo{author}{\bibfnamefont{A.}~\bibnamefont{Yacoby}},
  \bibinfo{journal}{Nat. Phys.} \textbf{\bibinfo{volume}{5}},
  \bibinfo{pages}{903} (\bibinfo{year}{2009}).

\bibitem[{\citenamefont{Gullans et~al.}(2010)\citenamefont{Gullans, Krich,
  Taylor, Bluhm, Halperin, Marcus, Stopa, Yacoby, and Lukin}}]{Gullans10}
\bibinfo{author}{\bibfnamefont{M.}~\bibnamefont{Gullans}},
  \bibinfo{author}{\bibfnamefont{J.~J.} \bibnamefont{Krich}},
  \bibinfo{author}{\bibfnamefont{J.~M.} \bibnamefont{Taylor}},
  \bibinfo{author}{\bibfnamefont{H.}~\bibnamefont{Bluhm}},
  \bibinfo{author}{\bibfnamefont{B.~I.} \bibnamefont{Halperin}},
  \bibinfo{author}{\bibfnamefont{C.~M.} \bibnamefont{Marcus}},
  \bibinfo{author}{\bibfnamefont{M.}~\bibnamefont{Stopa}},
  \bibinfo{author}{\bibfnamefont{A.}~\bibnamefont{Yacoby}}, \bibnamefont{and}
  \bibinfo{author}{\bibfnamefont{M.~D.} \bibnamefont{Lukin}},
  \bibinfo{journal}{Phys. Rev. Lett.} \textbf{\bibinfo{volume}{104}},
  \bibinfo{pages}{226807} (\bibinfo{year}{2010}).

\bibitem[{\citenamefont{Bluhm et~al.}(2010)\citenamefont{Bluhm, Foletti,
  Mahalu, Umansky, and Yacoby}}]{Bluhm10}
\bibinfo{author}{\bibfnamefont{H.}~\bibnamefont{Bluhm}},
  \bibinfo{author}{\bibfnamefont{S.}~\bibnamefont{Foletti}},
  \bibinfo{author}{\bibfnamefont{D.}~\bibnamefont{Mahalu}},
  \bibinfo{author}{\bibfnamefont{V.}~\bibnamefont{Umansky}}, \bibnamefont{and}
  \bibinfo{author}{\bibfnamefont{A.}~\bibnamefont{Yacoby}},
  \bibinfo{journal}{Phys. Rev. Lett.} \textbf{\bibinfo{volume}{105}},
  \bibinfo{pages}{216803} (\bibinfo{year}{2010}).

\bibitem[{\citenamefont{Zhang et~al.}(2010)\citenamefont{Zhang, Hu, Zhuang,
  You, and Liu}}]{Zhang10b}
\bibinfo{author}{\bibfnamefont{W.}~\bibnamefont{Zhang}},
  \bibinfo{author}{\bibfnamefont{J.-L.} \bibnamefont{Hu}},
  \bibinfo{author}{\bibfnamefont{J.}~\bibnamefont{Zhuang}},
  \bibinfo{author}{\bibfnamefont{J.~Q.} \bibnamefont{You}}, \bibnamefont{and}
  \bibinfo{author}{\bibfnamefont{R.-B.} \bibnamefont{Liu}},
  \bibinfo{journal}{Phys. Rev. B} \textbf{\bibinfo{volume}{82}},
  \bibinfo{pages}{045314} (\bibinfo{year}{2010}).

\bibitem[{\citenamefont{Gullans et~al.}()\citenamefont{Gullans, Krich, Taylor,
  Halperin, and Lukin}}]{Gullans13}
\bibinfo{author}{\bibfnamefont{M.}~\bibnamefont{Gullans}},
  \bibinfo{author}{\bibfnamefont{J.~J.} \bibnamefont{Krich}},
  \bibinfo{author}{\bibfnamefont{J.~M.} \bibnamefont{Taylor}},
  \bibinfo{author}{\bibfnamefont{B.~I.} \bibnamefont{Halperin}},
  \bibnamefont{and} \bibinfo{author}{\bibfnamefont{M.~D.} \bibnamefont{Lukin}},
  \eprint{arXiv:1212.6953v1 [cond-mat]}.

\bibitem[{\citenamefont{Taylor et~al.}(2007)\citenamefont{Taylor, Petta,
  Johnson, Yacoby, Marcus, and Lukin}}]{Taylor07}
\bibinfo{author}{\bibfnamefont{J.~M.} \bibnamefont{Taylor}},
  \bibinfo{author}{\bibfnamefont{J.~R.} \bibnamefont{Petta}},
  \bibinfo{author}{\bibfnamefont{A.~C.} \bibnamefont{Johnson}},
  \bibinfo{author}{\bibfnamefont{A.}~\bibnamefont{Yacoby}},
  \bibinfo{author}{\bibfnamefont{C.~M.} \bibnamefont{Marcus}},
  \bibnamefont{and} \bibinfo{author}{\bibfnamefont{M.~D.} \bibnamefont{Lukin}},
  \bibinfo{journal}{Phys. Rev. B} \textbf{\bibinfo{volume}{76}},
  \bibinfo{pages}{035315} (\bibinfo{year}{2007}).

\bibitem[{\citenamefont{Dobrovitski et~al.}(2006)\citenamefont{Dobrovitski,
  Taylor, and Lukin}}]{Dobrovitski06}
\bibinfo{author}{\bibfnamefont{V.~V.} \bibnamefont{Dobrovitski}},
  \bibinfo{author}{\bibfnamefont{J.~M.} \bibnamefont{Taylor}},
  \bibnamefont{and} \bibinfo{author}{\bibfnamefont{M.~D.} \bibnamefont{Lukin}},
  \bibinfo{journal}{Phys. Rev. B} \textbf{\bibinfo{volume}{73}},
  \bibinfo{pages}{245318} (\bibinfo{year}{2006}).

\bibitem[{\citenamefont{Misra and Sudarshan}(1977)}]{Misra77}
\bibinfo{author}{\bibfnamefont{B.}~\bibnamefont{Misra}} \bibnamefont{and}
  \bibinfo{author}{\bibfnamefont{E.~C.~G.} \bibnamefont{Sudarshan}},
  \bibinfo{journal}{J. Math. Phys.} \textbf{\bibinfo{volume}{18}},
  \bibinfo{pages}{756} (\bibinfo{year}{1977}).

\bibitem[{\citenamefont{Itano et~al.}(1990)\citenamefont{Itano, Heinzen,
  Bollinger, and Wineland}}]{Itano90}
\bibinfo{author}{\bibfnamefont{W.~M.} \bibnamefont{Itano}},
  \bibinfo{author}{\bibfnamefont{D.~J.} \bibnamefont{Heinzen}},
  \bibinfo{author}{\bibfnamefont{J.~J.} \bibnamefont{Bollinger}},
  \bibnamefont{and} \bibinfo{author}{\bibfnamefont{D.~J.}
  \bibnamefont{Wineland}}, \bibinfo{journal}{Phys. Rev. A}
  \textbf{\bibinfo{volume}{41}}, \bibinfo{pages}{2295} (\bibinfo{year}{1990}).

\bibitem[{\citenamefont{Klauser et~al.}(2008)\citenamefont{Klauser, Coish, and
  Loss}}]{Klauser08}
\bibinfo{author}{\bibfnamefont{D.}~\bibnamefont{Klauser}},
  \bibinfo{author}{\bibfnamefont{W.~A.} \bibnamefont{Coish}}, \bibnamefont{and}
  \bibinfo{author}{\bibfnamefont{D.}~\bibnamefont{Loss}},
  \bibinfo{journal}{Phys. Rev. B} \textbf{\bibinfo{volume}{78}},
  \bibinfo{pages}{205301} (\bibinfo{year}{2008}).

\bibitem[{\citenamefont{Stopa et~al.}(2010)\citenamefont{Stopa, Krich, and
  Yacoby}}]{Stopa10}
\bibinfo{author}{\bibfnamefont{M.}~\bibnamefont{Stopa}},
  \bibinfo{author}{\bibfnamefont{J.~J.} \bibnamefont{Krich}}, \bibnamefont{and}
  \bibinfo{author}{\bibfnamefont{A.}~\bibnamefont{Yacoby}},
  \bibinfo{journal}{Phys. Rev. B} \textbf{\bibinfo{volume}{81}},
  \bibinfo{pages}{041304(R)} (\bibinfo{year}{2010}).

\bibitem[{\citenamefont{Asshoff et~al.}(2011)\citenamefont{Asshoff, W\"ust,
  Merz, Litvinov, Gerthsen, Kalt, and Hetterich}}]{Asshoff11}
\bibinfo{author}{\bibfnamefont{P.}~\bibnamefont{Asshoff}},
  \bibinfo{author}{\bibfnamefont{G.}~\bibnamefont{W\"ust}},
  \bibinfo{author}{\bibfnamefont{A.}~\bibnamefont{Merz}},
  \bibinfo{author}{\bibfnamefont{D.}~\bibnamefont{Litvinov}},
  \bibinfo{author}{\bibfnamefont{D.}~\bibnamefont{Gerthsen}},
  \bibinfo{author}{\bibfnamefont{H.}~\bibnamefont{Kalt}}, \bibnamefont{and}
  \bibinfo{author}{\bibfnamefont{M.}~\bibnamefont{Hetterich}},
  \bibinfo{journal}{Phys. Rev. B} \textbf{\bibinfo{volume}{84}},
  \bibinfo{pages}{125302} (\bibinfo{year}{2011}).

\bibitem[{\citenamefont{Al-Hassanieh et~al.}(2006)\citenamefont{Al-Hassanieh,
  Dobrovitski, Dagotto, and Harmon}}]{Al-Hassanieh06}
\bibinfo{author}{\bibfnamefont{K.~A.} \bibnamefont{Al-Hassanieh}},
  \bibinfo{author}{\bibfnamefont{V.~V.} \bibnamefont{Dobrovitski}},
  \bibinfo{author}{\bibfnamefont{E.}~\bibnamefont{Dagotto}}, \bibnamefont{and}
  \bibinfo{author}{\bibfnamefont{B.~N.} \bibnamefont{Harmon}},
  \bibinfo{journal}{Phys. Rev. Lett.} \textbf{\bibinfo{volume}{97}},
  \bibinfo{pages}{037204} (\bibinfo{year}{2006}).

\bibitem[{\citenamefont{Zhang et~al.}(2007{\natexlab{c}})\citenamefont{Zhang,
  Konstantinidis, Al-Hassanieh, and Dobrovitski}}]{Zhang07r}
\bibinfo{author}{\bibfnamefont{W.}~\bibnamefont{Zhang}},
  \bibinfo{author}{\bibfnamefont{N.~P.} \bibnamefont{Konstantinidis}},
  \bibinfo{author}{\bibfnamefont{K.~A.} \bibnamefont{Al-Hassanieh}},
  \bibnamefont{and} \bibinfo{author}{\bibfnamefont{V.~V.}
  \bibnamefont{Dobrovitski}}, \bibinfo{journal}{J. Phys.: Condens. Matter}
  \textbf{\bibinfo{volume}{19}}, \bibinfo{pages}{083202}
  (\bibinfo{year}{2007}{\natexlab{c}}).

\end{thebibliography}

\end{document}